\documentclass[aps,amssymb,pra,amsmath,floatfix]{revtex4}
\usepackage{epsfig}
\usepackage{bm}
\usepackage{ulem}

\newcommand{\undertilde}[1]{\underset{\widetilde{\phantom{aa}}}{#1}}

\newcommand{\be}{\begin{equation}}
\newcommand{\ee}{\end{equation}}

\begin{document}

\title{Multi-site mean-field theory for cold bosonic atoms 
in optical lattices}

\author{T. McIntosh}
\affiliation{Department of Physics, Engineering Physics, and 
Astronomy, Queen's University, Kingston, ON K7L 3N6, Canada}

\author{P. Pisarski}
\affiliation{Department of Physics, Engineering Physics, and 
Astronomy, Queen's University, Kingston, ON K7L 3N6, Canada}

\author{R. J. Gooding}
\affiliation{Department of Physics, Engineering Physics, and 
Astronomy, Queen's University, Kingston, ON K7L 3N6, Canada}

\author{E. Zaremba}
\affiliation{Department of Physics, Engineering Physics, and 
Astronomy, Queen's University, Kingston, ON K7L 3N6, Canada}

\date{\today}

\begin{abstract}
We present a detailed derivation of a multi-site mean-field theory 
(MSMFT) used to describe the Mott-insulator to
superfluid transition of bosonic atoms in optical lattices. The
approach is based on partitioning the lattice into small clusters 
which are decoupled by means of a mean field approximation. This
approximation invokes local superfluid order parameters defined 
for each of the boundary sites of the cluster. The resulting 
MSMFT grand potential has a non-trivial topology as a function 
of the various order parameters. An understanding of
this topology provides two different criteria
for the determination of the Mott insulator
superfluid phase boundaries. We apply this formalism to 
$d$-dimensional hypercubic lattices in one, two and three dimensions,
and demonstrate the improvement in the estimation of the phase 
boundaries when MSMFT is utilized for increasingly larger clusters,
with the best quantitative agreement found for $d=3$. The MSMFT is 
then used to examine a linear dimer chain in which the on-site
energies within the dimer have an energy separation of
$\Delta$. This system has
a complicated phase diagram within the parameter space of the
model, with many distinct Mott phases separated by superfluid
regions. 
\end{abstract}

\pacs{
67.85.Hj, 
03.75.Lm,
03.75.Hh,
05.30.Jp
}

\maketitle
 
\section{Introduction}

The properties of both the pure and disordered Bose-Hubbard (BH) model 
were first elucidated in a remarkably insightful paper by Fisher 
{\it et al.}~\cite{Fisher89}. For the case of a pure (or homogeneous) system, it
was shown that a collection of interacting bosons on a lattice
undergoes a Mott insulator (MI) to superfluid (SF)
transition as the ratio $J/U$ is varied, where $J$ characterizes
the rate of hopping from one lattice site to another, and $U$ represents
the interaction between two bosons residing on a given site. The
insulating phase appears as lobes in the $\mu/U$ vs. $J/U$ plane,
where $\mu$ is the chemical potential, within which the average
occupancy per site takes on an integral value. Outside these
bounded regions, the system is a superfluid characterized by a nonzero
condensate order parameter, and the density varies continuously
with the system parameters.

The BH model became particularly germane with the advent of trapped 
Bose gases and Jaksch {\it et al.}~\cite{Jaksch98} proposed that
the model would provide a realistic theoretical description of
bosons residing in an optical lattice. They obtained estimates of
the parameters appearing in the BH model and argued that the model
is relevant to the physical systems experimentally accessible.
Subsequent experiments~\cite{Bloch} indeed showed that trapped 
Bose gases are
an ideal setting within which to study the theoretically predicted
MI-SF transition.

Many theoretical studies of the BH model followed the original
Fisher {\it et al.} paper~\cite{Fisher89} using a variety of 
theoretical methods
and approximations and the properties of the transition in one,
two and three dimensions are now well
established~\cite{Batrouni92,Freericks96,Kuhner98,Rousseau06,Sansone07,
Sansone08,Santos09, Teichmann09}. Apart from the considerable 
work done on the disordered BH
model~\cite{Fisher89,Batrouni92,Freericks96,Bissbort10,Pisarski11}, 
the model has also been extended to
superlattices~\cite{Rousseau06,Roth03,Buonsante04a,Buonsante04b,
Buonsante05a,Buonsante05b,Deng08,Hen09,Hen10,Chen10}, spinor
condensates~\cite{Svidzinsky03,Tsuchiya04,Rizzi05},
multi-component systems~\cite{Kuklov04,Isacsson05a,Iskin10,Chen10} 
and multiband situations~\cite{Isacsson05b,Larson09,Mering11}. 
With increasing complexity, 
the MI-SF phase diagram becomes increasingly richer in structure.

One of the most useful theoretical approaches for obtaining a
qualitative understanding of the MI-SF transition is mean-field
theory which was originally motivated by considering the
infinite-range hopping limit~\cite{Fisher89}. Subsequent reformulations of
mean-field theory~\cite{Sheshadri93,Vanoosten01} invoked the existence 
of a condensate order
parameter which was used to decouple the nonlocal hopping term of
the BH Hamiltonian. In its simplest form, which we refer to as the
site-decoupled mean-field theory (SDMFT), the decoupling leads to a
system Hamiltonian consisting of a sum 
of site Hamiltonians. The latter are effectively independent but
depend on the order parameter, in general site-dependent, which
can be thought of as a variational parameter. For a homogeneous
lattice, the ground state of the system is determined by
minimizing the system energy with respect to the
order parameter. If the energy is minimized for
a non-zero value of the order parameter, the system is in the SF
phase, otherwise it is in the MI phase. The phase boundaries
obtained using this approach are consistent with the results of
more sophisticated approaches. It can be shown~\cite{Sheshadri93} 
that SDMFT is 
equivalent to the alternative starting point based on the 
Gutzwiller ansatz for the ground-state wave
function~\cite{Rokhsar91,Krauth92}.

One of the limitations of the SDMFT is the neglect of inter-site
correlations which allow for fluctuations of various physical
variables. For example, in the MI phase of a homogeneous system,
the site occupancy is precisely integral, whereas in reality some
(albeit small) fluctuations in the site occupancy must occur.
These effects can be captured, at least to some extent, by 
dividing the system into clusters
of arbitrary size and using mean-field theory to decouple the
clusters. We refer to theories of this kind as multi-site
mean-field theories (MSMFT) and in this paper, explore this
approximation for various situations. This approach has been used
previously by Buonsante {\it et
al.}~\cite{Buonsante04a,Buonsante05a} to study the MI-SF transition
in superlattices where novel (loophole)
features emerge and in an investigation 
of the disorderd BH model \cite{Pisarski11}. It should be
noted that a two-site mean-field theory was also introduced
in~\cite{Jain04} using the so-called phase-space method.

The main purpose of this paper is to provide a detailed derivation 
of the MSMFT and to
explain how it can be used to systematically improve the quality
of the results for the MI-SF phase boundary. It should be
emphasized at the outset that the method cannot compete in a
quantitative sense with more sophisticated methods such as
quantum Monte Carlo 
(QMC)~\cite{Batrouni92,Rousseau06,Sansone07,Sansone08}, 
the density-matrix renormalization-group (DMRG)
method~\cite{Kuhner98,Rizzi05} and other theoretical techniques
~\cite{Freericks96,Santos09}.
However, the MSMFT method is relatively straightforward and allows
one to explore efficiently the dependence of its predictions on 
the various
parameters which define more complex physical models. In addition,
as stated earlier, it has the merit
of providing useful information about inter-site correlations that
are missed in the SDMFT.

Our paper is organized as follows. In Sec. \ref{1D} we derive the
MSMFT for the simplest case of a one-dimensional lattice where the
definition of clusters used in the multi-site decoupling 
is particularly straightforward. The method is extended in Sec.
\ref{2D} to higher dimensions where more freedom is available,
within certain limits, in defining the clusters which cover the
lattice. The MI-SF transition is determined by means of a grand
potential which is a function of the various superfluid order
parameters defined for the cluster. A detailed analysis of the
grand potential leads to two different criteria for the
determination of the phase boundaries.
Our results for the case of homogeneous lattices are
presented in Sec.~\ref{Homogeneous}. In Sec.~\ref{Diatomic
Results} we consider one-dimensional superlattices; the
simplest consists of alternating $A$ and $B$
sites which we refer to as a dimer chain. In this case,
the nonequivalence of the sites requires two
different order parameters for the description of the superfluid
phase. We also consider an example of a four-site superlattice where 
qualitative differences appear in the predictions of SDMFT and
MSMFT.

\section{Formalism: Multi-Site Mean-Field Theory}

\subsection{Derivation in One Dimension}
\label{1D}

Our work is based on the BH Hamiltonian~\cite{Jaksch98}
in the grand canonical ensemble. With the assumption of a single
orbital per site, this Hamiltonian is given by
\begin{equation}\label{GCBH}
\hat{\mathcal{K}}=\hat{\mathcal{H}}~-~\mu\hat{\mathcal {N}}=
 \sum_{i} \left( \varepsilon_i~-~\mu\right) \hat{n}_{i}
 +\frac{1}{2}\sum_i U_i \hat{n}_{i}(\hat{n}_{i}~-~1)
 - \sum_{ij} J_{ij} \hat{c}^\dagger_{i} \hat{c}_{j} ,
\end{equation}
where the index $i$ labels the sites of the optical lattice and
$\hat {c}_{i}^\dagger$ and $\hat {c}_{i}$ are site creation and 
annihilation operators (henceforth referred to as site
operators), respectively; the number operator for site $i$ is 
given by $\hat{n}_{i}=\hat {c}^\dagger_{i}\hat {c}_{i}$. The
system parameters include the on-site energies $\varepsilon_i$
at each lattice site,
the tunnelling energy $J_{ij}$ between sites $i$ and $j$, and the intra-site
interaction energy $U_i$. To a good approximation it is
sufficient to ignore interactions between bosons on different
sites and hopping between sites further apart than the
nearest-neighbour distance~\cite{Jaksch98,Zhou10}. Furthermore, we
will restrict our considerations to the case where the
interaction parameter has a common value $U$ for all sites and a
hopping parameter $J$ for all nearest-neighbour pairs.
Generalizations to more complex situations such as
superlattices~\cite{Buonsante04a,Buonsante04b,Buonsante05a} can 
be readily
accommodated in the MSMFT that we develop. The final parameter in
the BH Hamiltonian is the chemical potential $\mu$ which controls 
the number of particles in the system. Although extensions to 
finite temperatures are certainly feasible~\cite{Buonsante04b},
we will only consider the properties of the BH
Hamiltonian at zero temperature.
 
\begin{figure*}[!ht]
\centering \scalebox{0.7}
{\includegraphics{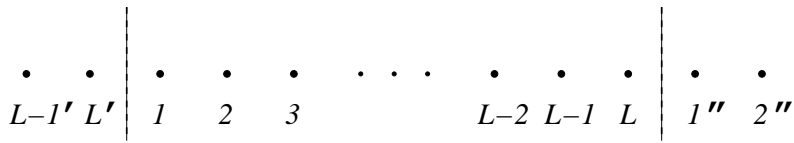}}
\caption{ The partitioning of a one-dimensional chain into linear 
clusters of length $L$. Within our mutli-site mean-field
formulation each cluster with open boundary conditions is treated 
exactly, and the inter-cluster couplings (generated by the hopping 
Hamiltonian) are treated
in a mean-field decoupling approximation.}
\label{fig:Lmers} 
\end{figure*}

The MSMFT~\cite{Buonsante04a} is most easily formulated for the 
example of a homogeneous one-dimensional chain
with nearest-neighbour hopping. The on-site energies
$\varepsilon_i$ can be set to zero and the interactions are
taken to be site independent. 
The first step in the derivation is to partition a
chain having $N_{s}$ sites into $N_c$ clusters each containing $L$
sites, so that
\begin{equation}
N_{s}=LN_{c}.
\end{equation}
We will refer to the cluster of length $L$ as an ``$L$-mer". A
schematic of the partitioning being considered is shown
in Fig.~\ref{fig:Lmers}. 
The hopping part of the Hamiltonian,
$\hat{\mathcal H}_{hop}=-J\sum_{\langle ij \rangle} 
\hat{c}^{\dagger}_{i}\hat{c}_{j}$, can then be written as 
\begin{equation}
\hat{\mathcal 
H}_{hop}=-J\sum_{j=0}^{N_{c}-1}\sum_{l=1}^{L-1}(\hat{c}^{\dagger}_{Lj
+l}\hat{c}_{Lj+l+1}+\hat{c}^{\dagger}_{Lj+l+1}\hat{c}_{Lj+l}
)-J\sum_{j=0}^{N_{c}-1}(\hat{c}^{\dagger}_{Lj+L}\hat{c}_{Lj+L+1}+\hat{c}
^{\dagger}_{Lj+L+1}\hat{c}_{Lj+L}).
\end{equation}
where we have isolated the terms with $l = L$ in the second sum
which couple sites between adjacent $L$-mers. This term will be
denoted by $\hat{\mathcal H}_{coup}$. We assume periodic
boundary conditions so that $\hat c_{LN_c+1}\equiv \hat c_{1}$.

The coupling between
$L$-mers can be eliminated by invoking the usual
argument~\cite{Sheshadri93,Vanoosten01}. We assume the existence
of a homogenous superfluid order parameter
\begin{equation}\label{orderparamdef}
\psi=\langle \hat{c}_{i} \rangle,
\end{equation}
and write
$\hat{c}_{i}=\psi  + (\hat{c}_{i}-\psi)$ for each of the operators
in $\hat{\mathcal H}_{coup}$.
Neglecting quadratic terms in the fluctuation $\delta \hat c_i =
\hat{c}_{i}-\psi$, we find
\begin{equation}
\hat{\mathcal 
H}_{coup}\simeq -J\sum_{j=0}^{N_{c}-1}\left[ 
( \hat{c}^{\dagger}_{Lj+L}+\hat{c}^{\dagger}_{L(j+1)+1} ) \psi + 
\left ( \hat{c}_{Lj+L} +\hat{c}_{L(j+1)+1}
\right ) \psi^* \right ] +2J N_{c}|\psi|^{2}
\end{equation}
We thus arrive at the cluster-decoupled Hamiltonian
\begin{equation}
\hat {\cal K} = \sum_{j=0}^{N_c-1} \hat {\cal K}_j^{MF},
\end{equation}
where $\hat {\cal K}_j^{MF}$ only depends on the 
site operators within
the $j$-th $L$-mer. Taking $j=0$, we have
\begin{equation}\label{KMF1D}
\hat{\mathcal K}_{0}^{MF}=\hat{\mathcal K}^{0}_{L}+\hat{\mathcal 
V}^{MF}_{L},
\end{equation}
where
\begin{equation}\label{K01D}
\hat{\mathcal 
K}^{0}_{L}=\frac{U}{2}\sum_{l=1}^{L}\hat{n}_{l}(\hat{n}_{l}-1)-\mu\sum_
{l=1}^{L}\hat{n}_{l}-J\sum_{l=1}^{L-1}(\hat{c}_{l}^{\dagger}\hat{c}_{l+1
}+\hat{c}_{l+1}^{\dagger}\hat{c}_{l}),
\end{equation}
and
\begin{equation}\label{VMF1D}
\hat{\mathcal 
V}^{MF}_{L}=-J\left [ \psi(\hat{c}^{\dagger}_{1}+
\hat{c}^{\dagger}_{L} ) + \psi^* \left ( \hat c_1 + 
\hat{c}_{L}\right)\right ] +2J|\psi|^{2}.
\end{equation}
The $\psi$-independent operator $\hat{\mathcal K}^{0}_{L}$ is the
Hamiltonian of a one-dimensional chain of length $L$ with open
ends. The different $L$-mers are independent physical systems but are 
effectively coupled by means of the order parameter appearing in
the mean-field perturbation $\hat{\mathcal V}^{MF}_{L}$.

It is clear from the form of (\ref{VMF1D}) that the
phase of the order parameter, $\psi = |\psi|\exp{(i\phi)}$, 
can be absorbed by a redefinition of the site operators:
$\exp{(-i\phi)}\hat c_l \to \hat
c_l$. The resulting Hamiltonian depends on $|\psi|$ and we
can therefore take the order parameter $\psi$ to be a real,
positive quantity. 
The ground state of the system described by a single homogeneous
order parameter can be determined by minimizing
$\hat{\mathcal K}_{0}^{MF}$ with respect to $\psi$ in th $L$-mer 
Fock space.

\subsection{MSMFT in Higher Dimensions}
\label{2D}

The extension of the MSMFT to higher dimensions is straightforward,
although some new elements appear due to the freedom available in
partitioning the lattice into clusters. This is illustrated in
Fig. 2(a) for the example of a two-dimensional square lattice. The
figure shows how the lattice can be covered using $L$-mers with
size $L = 1$, 2, 4, 5 and 6. It is clearly necessary that these
clusters cover the entire lattice without duplication. It is also
desirable that they have the same point-group symmetry as the
original lattice. The examples $L=1$, 4 and 9 satisfy this
latter criterion. The $L=5$ cluster has the point-group symmetry of the
square lattice, but its covering of the 2D plane introduces a
chirality not present in the homogeneous system.

\begin{figure*}[!ht]
\includegraphics[scale=0.55]{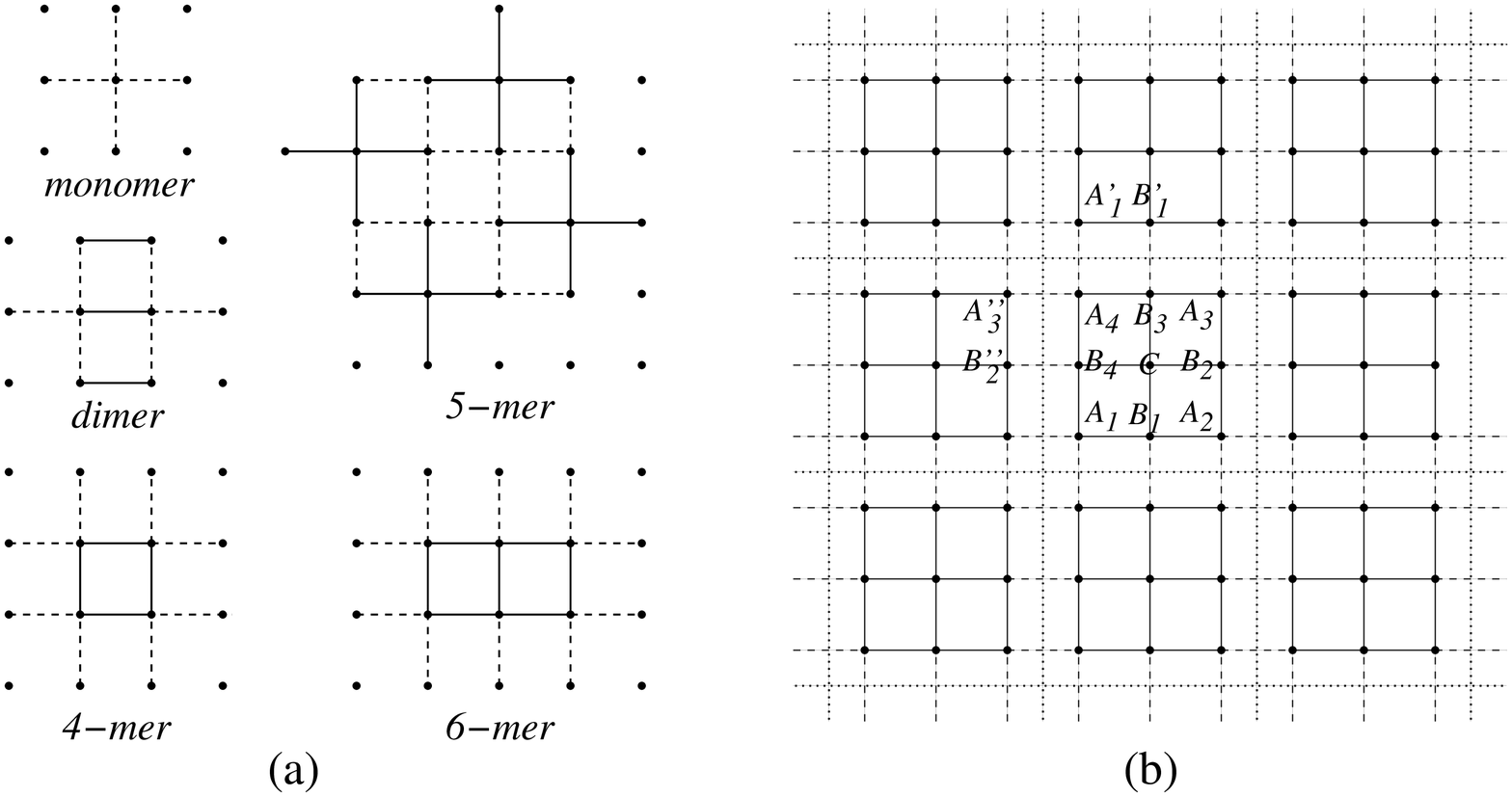}
\caption{
The partitioning of a two-dimensional square lattice using various 
repeated clusters. Shown in (a)
are single site monomers, dimers ($L=2$), and $L$-mers for
$L=4$, 5, 6.
Shown in (b) is  a portion of a 
two-dimensional square lattice that is partitioned into 
$3\times 3$ clusters. The solid lines connect sites in a given cluster, 
dotted lines show the division of the lattice into such clusters,
and the dashed lines show the inter-cluster couplings that are treated 
within mean-field theory. In each cluster there are three inequivalent 
sites: the central site $C$, which does not couple to neighbouring 
clusters, and corner and edge sites, which respectively are
labelled $A$ and $B$.}
\label{fig:3x3} 
\end{figure*}

Although we develop the MSMFT in its most general form, it will be 
useful to keep in mind the example of a two-dimensional square
lattice with the 3$\times$3 clusters shown in Fig. 2(b). The sites
in the lattice can be specified by ${\bf R} + \boldsymbol{\tau}$, where
${\bf R}$ is a cluster Bravais lattice vector and $\boldsymbol{\tau}$
defines the basis of sites within the cluster. For the example of
Fig. 2(b), $\boldsymbol{\tau}$ runs over the nine sites of the 3$\times$3
clusters. A site on the boundary of a cluster will be denoted by
$\alpha$ and is connected, via $J$, to one or more boundary sites on
adjacent clusters. The mean-field decoupling of the hopping term 
for the boundary site $\alpha$ in the $L$-mer and the boundary 
site $\beta'$ of the
adjoining $L'$-mer is achieved using the prescription
\begin{equation}\label{mfterm} 
\hat{c}^{\dagger}_{\alpha}\hat{c}_{\beta'}+\hat{c}^{\dagger}_{\beta'}\hat{c}_{\alpha}
\rightarrow
\psi_{\beta}\hat{c}^{\dagger}_{\alpha}
+ \psi_{\beta}^*\hat{c}_{\alpha}
+ \psi_{\alpha}\hat{c}^{\dagger}_{\beta'}
+ \psi_{\alpha}^*\hat{c}_{\beta'}
-\psi_\alpha^* \psi_{\beta}-\psi_\alpha \psi_{\beta}^*.
\end{equation}
In general, we allow for different order parameters for each of the
boundary sites. Since the $\beta'$ site in th $L'$-mer is related
to the $\beta$ site in the $L$-mer by
translational symmetry, we have denoted
the order parameter on this site as $\psi_\beta$.
(For example, the sites $A'_1$ and
$A_1$ in Fig. 2(b) must have the same order parameter by
translational symmetry.) 
For now, we have also allowed the
order parameters to be complex. Collecting all terms pertaining to
the $L$-mer, the mean-field decoupling leads to the $L$-mer
mean-field perturbation
\begin{equation}
\hat{\mathcal V}^{MF}_L(\{\psi_\alpha\}) \equiv -\sum_{\alpha\beta}
J_{\alpha\beta}
\left (  \hat{c}^{\dagger}_{\alpha}  \psi_\beta
+  \hat{c}_{\alpha}\psi_\beta^* 
- \psi_\alpha^* \psi_\beta \right ),
\label{MFP}
\end{equation}
where the sums extend over all boundary sites. To obtain this form
we have defined the symmetric matrix $\undertilde J$ with matrix
elements $J_{\alpha\beta} \equiv Jg_{\alpha\beta}$ where 
$g_{\alpha\beta}$ is the number of clusters to which the $L$-mer of 
interest is connected by a pair of $\alpha$ and $\beta$ sites in
the way described above. If such a pair of sites is not coupled,
$g_{\alpha\beta} =0$. We will refer to $\undertilde J$ as the {\it
connectivity} matrix, since it encapsulates the way in which the
different sites in the cluster are connected to one another
via the mean-field couplings~\cite{footnote}.

The perturbation in (\ref{MFP}) depends on the set of order parameters
$\{\psi_\alpha\}$ defined for the ensemble of boundary sites. 
The cluster Hamiltonian then takes the general form
\begin{equation}\label{KMF}
\hat{\mathcal K}^{MF}(\{\psi_{\alpha}\})=\hat{\mathcal 
K}^{0}_L+\hat{\mathcal V}^{MF}_L\left(\left\{\psi_{\alpha}\right\}\right),
\end{equation}
where the $\psi_\alpha$-independent part
\begin{equation}\label{K0}
\hat{\mathcal K}^{0}_L
=\frac{U}{2}\sum_{l=1}^{L}\hat{n}_{l}(\hat{n}_{l}-1)
+\sum_{l=1}^{L}(\varepsilon_l - \mu) \hat{n}_{l}-
J\sum_{\langle lm\rangle} \hat{c}_{l}^{\dagger}\hat{c}_m,
\end{equation}
is the sum of terms in the grand canonical 
Hamiltonian $\hat {\cal K}$ that depend only on the $L$-mer
variables. The final sum in (\ref{K0}) is restricted to
nearest-neighbour pairs within the $L$-mer. It is important to
note that this intra-cluster hopping leads to inter-site 
correlations even within the Mott insulating phase.

The MSMFT Hamiltonian depends
on various factors, such as the dimension and geometry of the
lattice, the shape and size of the clusters and the coordination of
sites in one cluster with those of its neighbours. 
The extent to which different order parameters are required
depends on the physical application and will be clarified by
example. However, we emphasize that these order parameters are
not prescribed but in general are determined by the solution of
the mean-field problem itself.

The cluster grand canonical Hamiltonian (\ref{KMF}) is Hermitian but 
it is not number-conserving due to the perturbation in (\ref{MFP}). As a
result, its eigenvectors must be determined in the $L$-mer Fock
space. We denote the state with the lowest
eigenvalue $\Omega_0(\{\psi_{\alpha}\})$ as $|\Psi_0\rangle$.
For a normalized state vector, the variation of
$\Omega_0(\{\psi_{\alpha}\})$ with respect to $\psi_\gamma^*$ is
given by
\begin{equation}
\frac{\partial \Omega_0(\{\psi_{\alpha}\})}{\partial \psi_\gamma^* } 
= \left \langle \Psi_0 \left | \frac{\partial \hat{\mathcal
V}^{MF}_L\left(\left\{\psi_{\alpha}\right\}\right)}{
\partial\psi_\gamma^* } \right |\Psi_0 \right \rangle = - \sum_\alpha
J_{\alpha\gamma} \langle \Psi_0 | \hat c_\alpha - \psi_\alpha
|\Psi_0\rangle.
\end{equation}
This implies that $\Omega_0(\{\psi_{\alpha}\})$ is stationary when
the order parameters satisfy
\begin{equation}
\langle \hat c_\gamma \rangle \equiv \langle \Psi_0(\{\bar
\psi_\alpha\}) |  \hat
c_\gamma |\Psi_0(\{\bar \psi_\alpha\})\rangle =
\bar{\psi}_\gamma
\label{self-consistency}
\end{equation}
for all boundary sites $\gamma$, where the values of the order
parameters at the stationary point are denoted by
$\bar{\psi}_\gamma$. We thus see that stationarity of the grand
potential is associated with the physically necessary condition
that the order parameters are determined self-consistently.
If more than one stationary point arises, the
physical state of the system is assumed to correspond to the stationary
point with the minimum value of $\Omega_0(\{\bar\psi_{\alpha}\})$.

Intuition might lead one to expect that a stationary point is
an extremum of the grand potential, but it is straightforward 
to show that it is {\it not}. Writing $\psi_\alpha =
\bar\psi_\alpha+\Delta\psi_\alpha$ and expanding (\ref{KMF}) about
a stationary point, we have 
\begin{equation}\label{KMF_stationary}
\hat{\mathcal K}^{MF}(\{\psi_{\alpha}\})=
\hat{\mathcal K}^{MF}(\{\bar \psi_{\alpha}\})
-\sum_{\alpha\beta} J_{\alpha\beta}
\left [ ( \hat{c}^{\dagger}_{\alpha} -\bar \psi_\alpha^*) \Delta \psi_\beta
+  \left ( \hat{c}_{\alpha}-\bar \psi_{\alpha} \right )\Delta \psi_\beta^* 
-\Delta \psi_\alpha^* \Delta\psi_\beta \right ].
\end{equation}
We denote the sum by $\Delta {\cal \hat V}^{MF}_L$ and consider it as a 
perturbation to the grand Hamiltonian
$\hat{\mathcal K}^{MF}(\{\bar \psi_{\alpha}\})$ at the stationary
point. The first order correction to the grand potential
$\Omega_0(\{\bar\psi_\alpha\})$ is
\begin{equation}
\Omega_0^{(1)}(\{\Delta\psi_\alpha\}) = \langle \Psi_0|\Delta
{\cal \hat V}^{MF}_L | \Psi_0 \rangle =
\sum_{\alpha\beta} J_{\alpha\beta}
\Delta \psi_\alpha^* \Delta\psi_\beta
\label{first_order}
\end{equation}
while the second order correction is
\begin{equation}
\Omega_0^{(2)}(\{\Delta\psi_\alpha\}) = \sum_{\nu\ne 0} 
\frac{|\langle \nu | \sum_{\alpha\beta} J_{\alpha\beta}
\left [ \hat{c}^{\dagger}_{\alpha} \Delta \psi_\beta
+  \hat{c}_{\alpha}\Delta \psi_\beta^* \right ] | 0 \rangle |^2}
{\Omega_0(\{\bar\psi_\alpha\})-\Omega_\nu(\{\bar\psi_\alpha\})},
\label{second_order}
\end{equation}
where the states $|\nu\rangle$ are eigenstates of $\hat{\mathcal
K}^{MF}(\{\bar \psi_{\alpha}\})$ with eigenvalues
$\Omega_\nu(\{\bar\psi_\alpha\})$.
$\Omega_0^{(2)}$ is negative definite and $\Omega_0^{(1)}$ can
always be made negative by choosing 
$\Delta \psi_\beta = - \Delta\psi_\alpha$ for some pair of
deviations with all others equal to zero. Thus, unlike the
situation for the case of a single order parameter, a stationary
point is not a local minimum; the value of
$\Omega_0(\{\psi_\alpha\})$ can always be made smaller than 
$\Omega_0(\{\bar \psi_\alpha\})$ by moving away from the stationary
point in some direction. For multiple order parameters,
the stationary point is, in general, a saddle point
and as a result, it cannot be located by
means of a variational principle. Below, we provide
criteria for identifying the emergence of a superfluid phase
from a Mott insulating phase without having to appeal to
a variational principle.

\subsection{Perturbative Treatment of the MI-SF Transition}
\label{phases}

The point $\{\psi_\alpha\} = \{0\}$ is always a stationary point
and it too is not an extremum in general. This point has special
significance in that it corresponds to the Mott insulating phase. 
Assuming the MI-SF transition to be continuous as a function of
the system parameters (for example, the hopping strength $J$), we
expect the stationary point to move continuously away from the
$\{\psi_\alpha\} = \{0\}$ point as the parameters are varied
beyond some critical values. This behaviour can be analyzed by
treating (\ref{MFP}) as a perturbation to ${\cal \hat K}^0_L$. Depending
on the values of the system parameters ($\mu$, $J$ and $U$), 
${\cal \hat K}^0_L$ has a ground state $|0\rangle$ containing 
$N$ particles. The range of parameters for which this is the case
defines what we refer to as $N$-domains in the multi-dimensional
parameter space. Each of these $N$-domains can, in principle, give
rise to a Mott phase with a certain number of particles per
cluster. These
regions are conventionally referred to as Mott lobes.

The first and second order corrections to the energy
$\Omega_0(\{0\})$ are obtained with the replacement $\bar\psi_\alpha
\to 0$ and $\Delta \psi_\alpha \to \psi_\alpha$ in
(\ref{first_order}) and (\ref{second_order}). We
have
\begin{equation}
\Omega_0^{(1)}(\{\psi_\alpha\}) = 
\sum_{\alpha\beta} J_{\alpha\beta}~
\psi_\alpha^* \psi_\beta
\label{first_order_0}
\end{equation}
and
\begin{equation}
\Omega_0^{(2)}(\{\psi_\alpha\}) = \sum_{\nu\ne 0} 
\frac{|\langle \nu | \sum_{\alpha\beta} J_{\alpha\beta}
\left [ \hat{c}^{\dagger}_{\alpha} \psi_\beta
+  \hat{c}_{\alpha} \psi_\beta^* \right ] | 0 \rangle |^2}
{\Omega_0(\{0\})  - \Omega_\nu(\{0\})},
\end{equation}
where in this case, the states $|\nu\rangle$ are eigenstates of $\hat
{\cal K}^0_L$ and are therefore number eigenstates. For the matrix
elements in the sum to be nonzero, the state $|\nu\rangle$ 
must have either $N - 1$ or $N+1$ particles. Defining the operator
\begin{equation}
\hat O_\alpha = \sum_\beta J_{\alpha\beta}~{\hat c_\beta},
\label{O-op}
\end{equation}
we see that the second order correction is given by
\begin{equation}
\Omega_0^{(2)}(\{\psi_\alpha\}) = \sum_{\alpha \beta} 
M_{\alpha\beta} ~\psi_\alpha^*\psi_\beta
\label{second_order_0}
\end{equation}
where the Hermitian matrix $\undertilde{M}$ is defined by
\begin{equation}
M_{\alpha\beta} = \sum_{\nu\ne 0} 
\frac{\langle 0 | \hat O_\alpha
|\nu\rangle \langle \nu| \hat O_\beta^\dagger |0\rangle
+ \langle 0 | \hat O_\beta^\dagger
|\nu\rangle \langle \nu| \hat O_\alpha |0\rangle}
{\Omega_0(\{0\})  - \Omega_\nu(\{0\})}.
\end{equation}
In obtaining this result we have used the fact that the operator
$\hat O_\alpha$ only has finite matrix elements between states
whose particle numbers differ by one.
In view of (\ref{O-op}), we see that 
\be
\undertilde{M} =
\undertilde{J}\undertilde{C}\undertilde{J}
\ee
where the elements of the Hermitian matrix $\undertilde{C}$ are
given by
\begin{equation}
C_{\alpha\beta} = \sum_{\nu\ne 0} 
\frac{\langle 0 | \hat c_\alpha
|\nu\rangle \langle \nu| \hat c_\beta^\dagger |0\rangle
+ \langle 0 | \hat c_\beta^\dagger
|\nu\rangle \langle \nu| \hat c_\alpha |0\rangle}
{\Omega_0(\{0\})  - \Omega_\nu(\{0\})}.
\label{C}
\end{equation}
Combining (\ref{first_order_0}) and (\ref{second_order_0}), we
have
\begin{equation}
\Omega_0(\{\psi_\alpha\}) = \Omega_0(\{0\}) + \sum_{\alpha\beta}
W_{\alpha\beta} ~\psi_\alpha^*\psi_\beta + \cdots,
\label{Omega_0}
\end{equation}
where 
\begin{equation}
\undertilde{W} \equiv
\undertilde{J}+\undertilde{M}=\undertilde{J} +
\undertilde{J}\undertilde{C}\undertilde{J}. 
\label{W_matrix}
\end{equation}
We will refer to the
Hermitian matrix $\undertilde{W}$ as the energy matrix.

If the value $\Omega_0(\{0\})$ is the minimum of all stationary
points, one is in the Mott insulating phase. The question then
arises as to whether the stability of the Mott phase can be
established independently of determining
$\Omega_0(\{\bar\psi_\alpha\})$ for all of these points.
Since $\Omega_0(\{0\})$ is not an extremum, it is not obvious 
{\it a priori} what
properties the energy matrix $\undertilde{W}$ must have in order
for the Mott phase to be stable. To address this query we
determine the eigenvectors and eigenvalues of $\undertilde{W}$:
\begin{equation} 
\undertilde{W} ~{\bf v}_i = \omega_i ~{\bf v}_i.
\label{W_eigen}
\end{equation}
Since $\undertilde{W}$ is Hermitian, the eigenvalues are real and
the eigenvectors can be chosen to be an orthonormal set. Expanding
the order-parameter vector $\boldsymbol{\psi} \equiv \{\psi_\alpha\}$ as
\begin{equation}
\boldsymbol{\psi} = \sum_i \xi_i ~{\bf v}_i,
\end{equation}
we see that
\begin{equation}
\Delta\Omega_0(\boldsymbol{\psi}) = \sum_i \omega_i |\xi_i|^2 .
\end{equation}
The eigenvectors ${\bf v}_i$ thus define directions in the 
order-parameter 
space along which the grand potential varies quadratically with
a curvature that is determined by the sign of $\omega_i$.
Furthermore, as one moves along a line in one of these directions, 
the grand potential is stationary with respect to displacements
away from the line. Thus the transition to the SF phase must be
accompanied by the appearance of a stationary point
with $\boldsymbol{\psi} \ne {\bf 0}$ along one
of these directions. The only scenario consistent with the
continuous movement of the stationary point away from
$\boldsymbol{\psi} = {\bf 0}$ is that one of the positive
eigenvalues passes through zero. The first positive eigenvalue
to do so establishes the criterion for the transition to the SF
phase. This is very analogous to the situation one finds with a 
single order parameter~\cite{Vanoosten01}.
We elaborate on these properties of the grand potential below,
and illustrate the general behaviour with a concrete numerical 
example in Sec.~\ref{example}.

We now denote the vector relevant to the transition as ${\bf
v}_1$. When $\omega_1$ passes through zero, the stationary point
moves continuously from $\boldsymbol{\psi} = {\bf 0}$ to a point
where the order parameter is non-zero. If higher order terms in
the perturbation expansion are retained, one expects the
grand potential to behave, in the vicinity of the transition, 
according to the Landau theory~\cite{Landau80} expression
\begin{equation}
\Delta\Omega_0(\boldsymbol{\psi}) = \frac{r}{2} |\xi_1|^2 +
\frac{u}{4}
|\xi_1|^4,
\label{eq:Omega_GL}
\end{equation}
where $r=2\omega_1$, and we assume $u$ to be
positive. For small and negative $\omega_1$, this leads to the
stationary order parameter
\begin{equation}
\boldsymbol{\psi} = \sqrt{\frac{|r|}{u}} {\bf v}_1.
\label{eq:psi_mf}
\end{equation}
Since the vector ${\bf v}_1$ can be chosen to be a real, one sees
that, in the vicinity of the phase boundary, the order parameter
is real. In other words, there are no phase differences (apart
from a possible sign) between the order parameter components. In
fact, in all the examples we have studied, the components of the
order parameter vector all have the {\it same} sign, which we
take to be positive. One might argue that this is the expected
behaviour, since relative sign differences in the components would
lead to more rapid spatial variations of the order parameter with
a resultant higher kinetic energy. However, as stated earlier, the
grand potential is not an extremum at the stationary point
and can take on lower
values if one moves away from the stationary point in some
direction ${\bf v}_i$. Since this vector is orthogonal to ${\bf
v}_1$, it must necessarily have components with different signs. 
One therefore
cannot argue that the components of the physically relevant
direction (${\bf v}_1$) must have the same sign on the basis of 
energy considerations.
Henceforth, we assume that the components of the order parameter
are real. This assumption is supported by all self-consistent
solutions of (\ref{KMF}) that we have obtained in the SF phase.

Referring to Eq.~(\ref{W_matrix}), we see that the eigenvalue
spectrum of $\undertilde{W}$ is closely related to that of the
connectivity matrix $\undertilde{J}$. In fact, for $J \to 0$, the
eigenvectors and eigenvalues of these two matrices coincide.
The eigenvectors of $\undertilde{J}$ are defined by
\begin{equation}\label{eq:Jeig}
\undertilde{J} {\bf u}_i = \lambda_i {\bf u}_i.
\end{equation}
The eigenvectors with zero eigenvalue play a special role in that
they are simultaneously zero-eigenvalue eigenvectors of 
$\undertilde{W}$. These vectors are independent of the
magnitude of $J$  and other system parameters, and are simply
determined by the structure of $\undertilde{J}$.
The eigenvectors of $\undertilde{W}$ with non-zero eigenvalues 
can be determined by using ${\bf u}_i$ as the basis vectors. 
Then, in what we call the $J$-representation, we
have
\begin{equation}
W_{ij}^{(J)} = \lambda_i\delta_{ij} + \lambda_i C_{ij}^{(J)}
\lambda_j.
\end{equation}
If we order the $\lambda_i = 0$ eigenvectors as $i=1,\dots,m$ and 
the remaining $n$ eigenvectors as $i=m+1,\dots,m+n$, the non-zero
eigenvectors of $\undertilde{W}$ are found in the
$n$-dimensional subspace spanned by the eigenvectors of
$\undertilde{J}$ with non-zero eigenvalues, and are determined by
\begin{equation}\label{eq:Wred-eig}
{\rm det}\left ( W_{n\times n}^{(J)} - \omega I_{n\times n} \right ) =
0.
\end{equation}
Here we have defined the $n\times n$ matrix 
\be
W_{n\times n}^{(J)} =D_{n\times n} + D_{n\times n}C_{n\times
n}^{(J)}D_{n\times n}
\ee
where $D_{n\times n}$ is the diagonal matrix
\begin{equation}
D_{n\times n} = \left ( \begin{array}{cccc}
\lambda_{m+1} & 0 & \cdots & 0\\
0 & \lambda_{m+2} &  & \vdots \\
\vdots &  &  \ddots  & 0 \\
0 & \cdots &  0 &\lambda_{m+n}\\
\end{array} \right ),
\label{Dnxn}
\end{equation}
and
\begin{equation}
C_{n\times n}^{(J)} = \left ( \begin{array}{ccc}
C_{m+1,m+1}^{(J)}
&\cdots & C_{m+1,m+n}^{(J)}\\
\vdots & \ddots & \vdots \\
C_{m+n,m+1}^{(J)} &\cdots & C_{m+n,m+n}^{(J)}\\
\end{array} \right ).
\end{equation}
The MI-SF transition is located by following the eigenvalues 
determined by (\ref{eq:Wred-eig}) as a function of a system
parameter,
such as $J$. As discussed above, the positive eigenvalue
$\omega_1$ which first passes through zero determines the phase
boundary. In view of (\ref{eq:Wred-eig}), the condition for this
to happen is ${\rm det} (W_{n\times n}^{(J)} ) = 0$.

Although the eigenvalue problem  in 
(\ref{eq:Wred-eig}) eliminates the eigenvectors with zero eigenvalues,
in practice it is more straightforward to determine all the
eigenvalues and eigenvectors using (\ref{W_eigen}). When the
$\omega_1$ eigenvalue is identified and found to pass through zero,
the corresponding eigenvector directly determines the
relative magnitude of the order parameter components just as one
enters the SF phase. The main advantage of the
$J$-representation is that is reveals the mathematical structure
of the $\undertilde{W}$ matrix and facilitates some of the
formal developments that follow.

An alternate approach to that of following the $\undertilde{W}$
eigenvalues is based on constructing a stability criterion for 
the Mott phase~\cite{Buonsante05a}. It too is based on a
perturbative analysis and
can be derived as follows. To first order in perturbation theory,
one finds that
\begin{equation}
\langle \hat c_\alpha \rangle^{(1)} = \sum_\beta S_{\alpha\beta}~\psi_\beta,
\label{order_param}
\end{equation}
where the matrix $\undertilde{S}$, which we refer to as the 
stability matrix, is defined in terms of matrices introduced 
earlier, namely
\begin{equation}
\undertilde{S} = - \undertilde{C} \undertilde{J}.
\label{S}
\end{equation}
We note that this matrix is {\it not} Hermitian.
Since the left-hand side of Eq.~(\ref{order_param}) is the
perturbative estimate of $\psi_\alpha$, this equation is suggestive 
of an iterative scheme defined 
by the linear map
\begin{equation}
\psi_\alpha^{(k+1)} = \sum_\beta
S_{\alpha\beta}\psi_\beta^{(k)},
\label{map}
\end{equation}
where $k = 0,\,1,\,2,\dots$ can be thought of as an iteration 
index with $k\rightarrow\infty$ identifying the self-consistent 
solution.
Defining the eigenvectors of $\undertilde{S}$ by
\begin{equation}
\undertilde{S}{\bf z}_i = \sigma_i {\bf z}_i,
\label{S_eigen}
\end{equation}
we see that the
iterative sequence converges to the Mott insulating  
$\boldsymbol{\psi} = {\bf 0}$ 
fixed point if the absolute values of the eigenvalues
$\sigma_i$ are all less than one.  Therefore, this procedure
provides an operational definition of the Mott phase in terms of
the eigenvalues of the stability matrix $\undertilde{S}$.

The definition of the stability matrix in (\ref{S}) shows that 
the eigenvectors
of $\undertilde{J}$ with zero eigenvalues are also zero-eigenvalue
eigenvectors of $\undertilde{S}$, that is, ${\bf z}_i = {\bf u}_i$
for $i = 1,\dots,m$, with $\sigma_i = 0$. In the
$J$-representation, the stability matrix thus has the block structure
\begin{equation}
\undertilde{S}^{(J)} = \left ( \begin{array}{cc}
O_{m\times m} & S_{m\times n}^{(J)}\\
O_{n\times m} & S_{n\times n}^{(J)} \\
\end{array} \right ),
\label{S_J}
\end{equation}
where the null matrices ($O$) have the dimensions indicated,
\begin{equation}
S_{n\times n}^{(J)} = - C_{n\times n}^{(J)}D_{n\times n}
\label{S_J_nxn}
\end{equation}
and
\begin{equation}
S_{m\times n}^{(J)} = \left ( \begin{array}{ccc}
-C_{1,m+1}^{(J)}\lambda_{m+1} &\cdots & -C_{1,m+n}^{(J)}\lambda_{m+n}\\
\vdots & \ddots & \vdots \\
-C_{m,m+1}^{(J)}\lambda_{m+1} &\cdots & -C_{m,m+n}^{(J)}\lambda_{m+n}\\
\end{array} \right ).
\label{S_J_mxn}
\end{equation}
It is clear from (\ref{S_J}) that the $\sigma = 0$ eigenvalue is
$m$-fold degenerate and that the remaining eigenvalues of
$\undertilde{S}$ are determined by the equation
\begin{equation}\label{eq:Sred-eig}
{\rm det}\left ( S_{n\times n}^{(J)} - \sigma I_{n\times n} \right ) =
0.
\end{equation}
If these latter eigenvalues all have a magnitude less than one,
then $\boldsymbol{\psi}={\bf 0}$ is a fixed point of the map
defined in (\ref{map}), and the Mott phase is stable.

The complete set of eigenvectors of $\undertilde{S}$ consist of
the $m$ zero-eigenvalue eigenvectors of $\undertilde{J}$ together with
the remaining $n$ eigenvectors determined by considering the
$n$-dimensional eigenvalue problem
\begin{equation}
S_{n\times n}^{(J)} ~{\bf y}_i = \sigma_i ~{\bf y}_i.
\end{equation}
In view of (\ref{S_J_nxn}), this equation is equivalent to the
generalized eigenvalue problem
\begin{equation}
C_{n\times n}^{(J)} {\bf y}_i' = -\sigma_i
D_{n\times n}^{-1}{\bf y}_i',
\end{equation}
where ${\bf y}_i' \equiv D_{n\times n} {\bf y}_i$. Since
$C_{n\times n}^{(J)}$ is Hermitian and $D_{n\times n}^{-1}$ (the
inverse of (\ref{Dnxn})) is a
real diagonal matrix, it is easy to
show that the eigenvalues $\sigma_i$ are real. In addition, for
distinct eigenvalues, we have the orthogonality relation
$(D_{n\times n}~ {\bf y}_i)^T {\bf y}_j = 0$. Once the vectors
${\bf y}_i$ have been determined, the $\undertilde{S}^{(J)}$
eigenvectors for $i=m+1,\dots,m+n$ are given by
\begin{equation}
{\bf z}_i = \left ( \begin{array}{c}
{\bf x}_i\\
{\bf y}_i
\end{array} \right ),
\label{S_eigenvec}
\end{equation}
with ${\bf x}_i = \sigma_i^{-1} S_{m\times n}^{(J)} {\bf y}_i$.

We now establish the connection between the two different criteria 
derived above. 
The energy and stability matrices are related by
\begin{equation}\label{eq:WJIS}
\undertilde{W} = \undertilde{J}(\undertilde{I}-\undertilde{S}).
\end{equation}
An explicit expression for $\undertilde{S}$ cannot be obtained from
this equation since $\undertilde{J}$ does not in general have an
inverse. However, within
the $J$-representation, we have the more useful relation
\begin{equation}
W_{n\times n}^{(J)} = D_{n\times n} (I_{n\times n}-S_{n\times
n}^{(J)}).
\end{equation}
Since $D_{n\times n} $ has an inverse, we can write
\begin{equation}
S_{n\times n}^{(J)} = I_{n\times n}-D_{n\times n}^{-1}W_{n\times
n}^{(J)}.
\label{eq:S-W}
\end{equation}
From this we see that an eigenvector
$(q_{m+1},\cdots,q_{m+n})^T$ of $W_{n\times n}^{(J)}$
with zero eigenvalue is an eigenvector of $S_{n\times n}^{(J)}$
with eigenvalue one. This particular zero eigenvalue of the
$\undertilde{W}$ matrix is obtained at
the critical values of the system parameters and the
corresponding eigenvector, previously
denoted by ${\bf v}_1$, is given by ${\bf v}_1=
\sum_{i=m+1}^{m+n} q_i {\bf u}_i$. Using the structure of the
$\undertilde{S}^{(J)}$ eigenvectors in (\ref{S_eigenvec}), we 
thus see that 
\begin{equation}\label{eq:transition}
\undertilde{S} {\bf v}_1 = {\bf v}_1,\,\,  {\rm when}\,\,  \undertilde{W}
{\bf v}_1 = 0.
\end{equation}
We have thus proved that the energy
and stability criteria for the transition to the SF phase are
equivalent and can therefore be used interchangeably.

\subsection{Self-Consistent Superfluid Solutions}

The previous subsection outlined two different algorithms for 
identifying the parameter values of the Bose-Hubbard
model for which the MI phase is stable, and 
an example of the application of these methods is presented in 
the next subsection. When the MI phase loses stability,
the $T=0$ ground state of the Bose-Hubbard model corresponds to
a superfluid state. The order parameter is then determined by
finding the ground state solution of (\ref{KMF}) which satisfies
the self-consistency condition given by (\ref{self-consistency}).

It is a simple matter to obtain such self-consistent solutions
numerically. The iterative map
\be
\psi_\gamma^{(k+1)} = \langle \Psi_0(\{\psi_\alpha^{(k)}\}) | \hat
c_\gamma | \Psi_0(\{\psi_\alpha^{(k)}\}) \rangle
\label{eq:iterative_map}
\ee
provides updated order parameter components in terms of their
current values. We find that any initial guess of the order
parameters, provided that they all have the same sign, converges
to a unique self-consistent solution without the need for
special mixing \cite{mixing} or more advanced techniques. 
This iterative method was used extensively in a recent
paper~\cite{Pisarski11} on disordered systems where a large
number of sites with different order parameters are required.
For homogeneous systems with only one order parameter, Brent's 
algorithm for root finding is particularly successful~\cite{Brent}.

\subsection{Example Calculation}
\label{example}

\begin{figure}[!h]
\centering \scalebox{0.6}
{\includegraphics{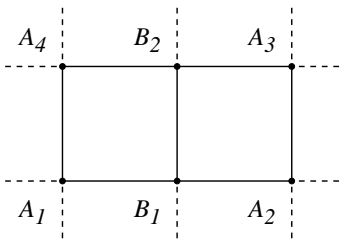}}
\caption{ The partitioning of the two-dimensional square lattice
into $3\times 2$ clusters. The cluster consists of inequivalent
corner ($A$) and edge ($B$) sites.}
\label{fig:3x2} 
\end{figure}

We now illustrate the calculation of the phase boundary and 
superfluid order parameters using the methods outlined in the 
previous two subsections. 
It is instructive to consider the 2D square lattice partitioned
into $3\times 2$ clusters with the site labelling 
indicated in Fig.~\ref{fig:3x2}. For this $L$-mer,
there are six boundary sites: the four corner ($A$) sites 
and the two edge ($B$) sites.
The choice of this particular cluster is useful for a number of
reasons. First, it demonstrates that meaningful physical results
can be obtained even though the cluster does not have the symmetry 
of the two-dimensional square lattice. Second, it is a
relatively simple example with several order parameters which
illustrates the general formalism and the structure of the
solutions. And lastly, it has obvious symmetries which allow us
to reduce the number of independent order parameters; here there
end up being two, one for the corner sites and one for the edge
sites. Thus, for demonstration purposes this is an ideal 
cluster to examine.

The calculation of the MI-SF phase boundary can be approached in
terms of the energy or stability criterion, each of which
involves the connectivity matrix $\undertilde{J}$.
For the site labelling shown in Fig.~\ref{fig:3x2} and the 
order parameter specified as $\boldsymbol{\psi} =
(\psi_{A_1},\psi_{A_2},\psi_{A_3},\psi_{A_4},\psi_{B_1},\psi_{B_2})^T$,
the connectivity matrix is given by
\begin{equation}
\undertilde{J} = \left ( \begin{array}{cccccc}
0 & J & 0 & J & 0 & 0\\
J & 0 & J & 0 & 0 & 0\\
0 & J & 0 & J & 0 & 0\\
J & 0 & J & 0 & 0 & 0\\
0 & 0 & 0 & 0 & 0 & J\\
0 & 0 & 0 & 0 & J & 0\\
\end{array} \right ).
\end{equation}
We observe that the matrix is block diagonal reflecting the fact that
corner sites are coupled to corner sites, and edge sites are coupled to
edge sites. Following the notation of (\ref{eq:Jeig}), the 
$\undertilde{J}$ matrix has two $\lambda = 0$
eigenvalues, and the non-zero eigenvalues are $\pm J$ and $\pm 2J$.
The energy (\ref{W_matrix}) and stability (\ref{S}) matrices
also depend on the $\undertilde{C}$ matrix in
(\ref{C}). The evaluation of this matrix requires the eigenstates 
$|\nu \rangle$ and corresponding eigenvalues $\Omega_\nu(\{0\})$ 
of the cluster Hamiltonian in (\ref{K0}) with on-site energies
$\varepsilon_l = 0$. We present details of the
evaluation of these states in an Appendix.

\begin{figure*}
\unitlength1cm
\begin{minipage}[t]{8cm}
\begin{picture}(6,12)
{\includegraphics[scale=0.75]{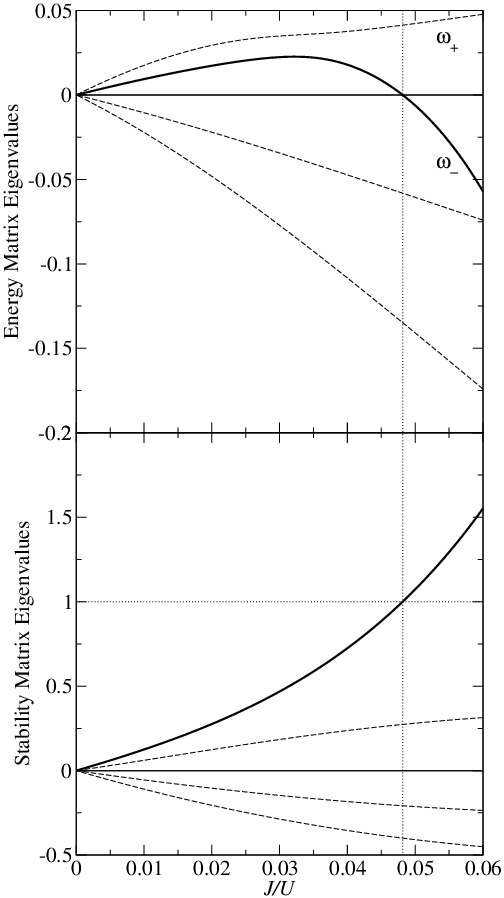}}
\end{picture}\par
\caption{The non-zero eigenvalues of the 
energy (top panel, in units of $U$) and stability (bottom panel)
matrices for a 
homogeneous two-dimensional square lattice using a $3\times 2$ 
cluster vs. $J/U$. The chemical potential is $\mu/U = 0.4$. One
eigenvalue of the energy matrix crosses zero at
$J/U=0.04815$, the same position at which one eigenvalue of
the stability matrix
first attains a magnitude of one (the solid black curve in each 
figure). The superfluid phase is the $T=0$ ground state 
for $J/U > 0.04815$. }
\label{fig:3x2eigval}
\end{minipage}
\hfill
\begin{minipage}[t]{8cm}
\begin{picture}(6,12)
{\includegraphics[scale=0.75]{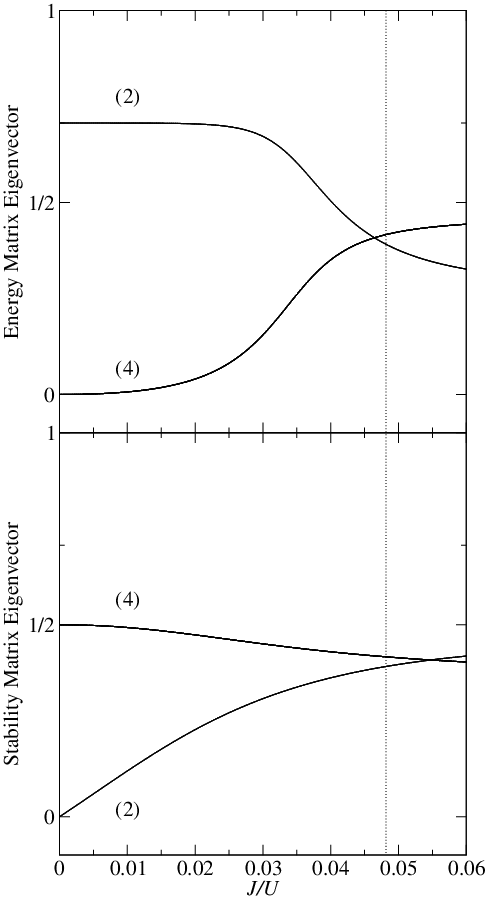}}
\end{picture}\par
\caption{The normalized eigenvector components
corresponding to the energy (top) and stability (bottom)
eigenvalues shown by the the solid black curves in 
Fig.~\ref{fig:3x2eigval} vs. $J/U$. The bracketed numbers 
give the number of components having the indicated values,
and correspond to the corner sites (multiplicity 4) and 
edge sites (multiplicity 2) of the $3\times2$ cluster. Note that 
at the transition these eigenvectors are the same, as required 
by (\protect\ref{eq:transition}).}
\label{fig:3x2eigvec}
\end{minipage}

\end{figure*}

As a concrete numerical example, we choose a chemical potential 
of $\mu/U=0.4$ which, for small $J/U$, places the system in the 
first Mott lobe with $\langle {\hat n} \rangle = 1$.
The eigenvalues of the energy matrix are obtained by solving
(\ref{W_eigen}), whereas the eigenvalues of the stability matrix
are obtained from (\ref{S_eigen}). The four non-zero eigenvalues
of these two matrices
are shown in Fig.~\ref{fig:3x2eigval} as a function of $J/U$. We
observe that one eigenvalue of the energy matrix associated
with the eigenvector ${\bf v}_1$ increases from
zero to positive values, and then passes through zero at the critical 
hopping $J_c/U = 0.04815$ which locates the MI-SF phase boundary
for this value of $\mu/U$. At this same hopping one
finds that one of the eigenvalues of the stability matrix first 
attains a magnitude of one, consistent with the discussion
following (\ref{eq:S-W}). Figure~\ref{fig:3x2eigvec} displays the 
variation of the eigenvector components of these two specific 
eigenstates,
namely, the eigenvector ${\bf v}_1$ of (\ref{W_eigen}) and the
eigenvector ${\bf z}_1$ of (\ref{S_eigen}). Both of these
eigenvectors have the form
$(\psi_A,\psi_A,\psi_A,\psi_A,\psi_B,\psi_B)^T$.
One sees that these two 
eigenvectors are the same (${\bf z_1} = {\bf v}_1$) at
the MI-SF phase boundary and that (\ref{eq:transition}) is
indeed satisfied.

\begin{figure}[!ht]
\centering{\includegraphics{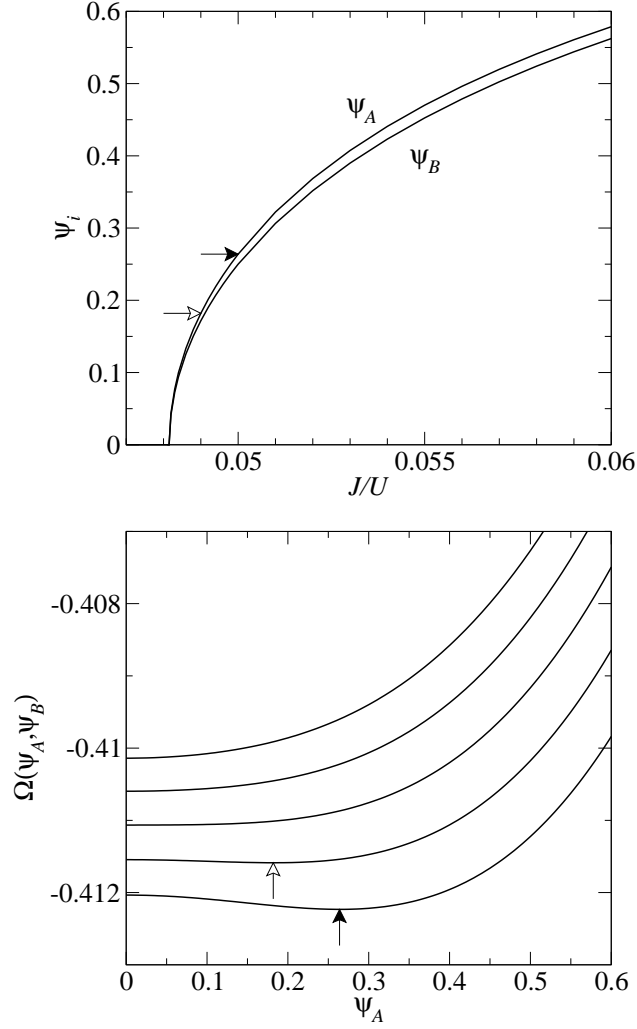}}
\caption{The upper figure shows the self-consistently determined 
order parameters
corresponding to the corner ($\psi_A$) and edge ($\psi_B$) sites
of the $3\times 2$ cluster as a function of $J/U$. The arrows
indicate the values of $\psi_A$ at $J/U = 0.049$ and 0.05.
The lower figure shows the grand
potential as a function of $\psi_A$ for the fixed ratio
$\psi_B/\psi_A = 0.9396$ found at the phase boundary at $J/U =
0.04815$. The curves from top to bottom are for values of 
$J/U$ from
0.046 to 0.050 in steps of 0.001. In the SF phase, the grand
potential shows minima at positions which are very
close to the values of $\psi_A$ found from the self-consistent 
calculations (indicated by the arrows). These results
were obtained for $\mu/U = 0.4$.
}
\label{fig:3x2psi}
\end{figure}

Although we allowed for a six-component order parameter in the
above analysis, we find that the relevant eigenvectors 
${\bf v}_1$ and ${\bf z}_1$ only have two independent components
corresponding to the two inequivalent sites in the $3\times 2$
cluster. One, $\psi_A$, is associated with the four corner sites
and the other, $\psi_B$, is associated with the two edge sites.
As can be seen in Fig.~\ref{fig:3x2eigvec}, these two values 
are almost equal for $J/U$ close to the MI-SF transition. 
Even though
the cluster does not mirror the symmetry of the square lattice,
the violation of homogeneity is relatively weak. This behaviour
persists into the SF phase. Fig.~\ref{fig:3x2psi} shows that the
self-consistently determined order parameters are close to each
other over the range of hoppings indicated. The variation shown
in Fig.~\ref{fig:3x2psi} is consistent with the mean-field 
prediction in (\ref{eq:psi_mf}), as is the ratio of the two 
components. Also shown in this figure is the grand potential
$\Omega_0(\{\psi_\alpha\})$ for the order parameter
$\boldsymbol{\psi} = (\psi_A,\psi_A,\psi_A,\psi_A,\psi_B,\psi_B)^T$ 
as a function of $\psi_A$,
where the ratio $\psi_B/\psi_A$ is fixed and taken from
Fig.~\ref{fig:3x2eigvec} at $J/U = 0.04815$. In the SF phase
the grand potential shows minima at points which are very close
to the order parameters determined self-consistently according
to (\ref{self-consistency}). Thus, near the phase boundary,
the grand potential is indeed given to a good approximation by
the Landau expansion in (\ref{eq:Omega_GL}). 

In view of the fact that the relevant eigenvector ${\bf v}_1$ only 
has two independent order parameter components, namely $\psi_A$ and
$\psi_B$, it is of interest to see to what extent the
calculations can be simplified by assuming an order parameter
having the form $\boldsymbol{\psi} = 
(\psi_A,\psi_A,\psi_A,\psi_A,\psi_B,\psi_B)^T$.
With this assumed form, the eigenvalue problem in
(\ref{W_eigen}) reduces to
\begin{equation}
\undertilde{W}^{\rm red} \boldsymbol{\psi}^{\rm red} \equiv \left ( \begin{array}{cc}
W_{AA} & W_{AB}\\
W_{BA} & W_{BB}\\ \end{array} \right )
\left ( \begin{array}{c}
\psi_A\\ \psi_B\\ \end{array} \right ) = \omega
\left ( \begin{array}{cc}
N_A & 0\\
0 & N_B\\ \end{array} \right )
\left ( \begin{array}{c}
\psi_A\\ \psi_B\\ \end{array} \right ),
\label{W_red}
\end{equation}
where $N_A$ and $N_B$ are the number of occurrences of $\psi_A$
and $\psi_B$, respectively, in the order parameter vector. For
this specific example, $N_A=4$ and $N_B=2$. The reduced matrix
elements in (\ref{W_red}) are defined as
\begin{equation}
W_{AA} = \sum_{i\in{A}}\sum_{j\in{A}} W_{ij},\quad
W_{AB} = \sum_{i\in{A}}\sum_{j\in{B}} W_{ij},\quad
W_{BA} = \sum_{i\in{B}}\sum_{j\in{A}} W_{ij},\quad
W_{BB} = \sum_{i\in{B}}\sum_{j\in{B}} W_{ij}.
\label{W_red_elem}
\end{equation}
By the same token,
the expansion of the grand potential in (\ref{Omega_0}) is given
by
\begin{equation}
\Omega_0(\psi_A,\psi_B) = \Omega_0(0,0) + 
\left ( \begin{array}{cc}
\psi_A & \psi_B\\ \end{array} \right ) 
\left ( \begin{array}{cc}
W_{AA} & W_{AB}\\
W_{BA} & W_{BB}\\ \end{array} \right )
\left ( \begin{array}{c}
\psi_A\\ \psi_B\\ \end{array} \right ).
\label{Omega_0_reduced}
\end{equation}
We note that the eigenvalues of the matrix $\undertilde{W}^{\rm red}$ 
appearing in this expression do {\it not} in general yield the desired
$\undertilde{W}$ eigenvalues which must be obtained from the
generalized eigenvalue problem in (\ref{W_red}).

Performing the sums in (\ref{W_red_elem}), we obtain the explicit
expressions
\begin{equation}
W_{AA} = z_AN_A J + z_A^2 J^2 \sum_{\nu \ne 0}
\frac{|\langle \nu| 
\sum_{i\in{A}}(\hat{a}_{i}^{\dagger}+\hat{a}_{i})|0\rangle|^{2}}
{\Omega_0(\{0\})-\Omega_\nu(\{0\})}~,
\label{W_AA}
\end{equation}
\begin{equation}
W_{BB} = z_BN_B J + z_B^2 J^2 \sum_{\nu \ne 0}
\frac{|\langle \nu| 
\sum_{i\in{B}}(\hat{b}_{i}^{\dagger}+\hat{b}_{i})|0\rangle|^{2}}{
\Omega_{0}(\{0\})-\Omega_\nu(\{0\})}~,
\label{W_BB}
\end{equation}
and
\begin{equation}
W_{AB} = z_A z_B J^2 \sum_{\nu \ne 0}
\frac{\langle 0| 
\sum_{i\in{A}}(\hat a_i^\dagger + \hat{a}_{i})|\nu\rangle\langle \nu |
\sum_{j\in{B}}(\hat b_i^\dagger + \hat{b}_{j})|0\rangle}
{ \Omega_{0}(\{0\})-\Omega_\nu(\{0\})}~.
\label{W_AB}
\end{equation}
Here we have introduced the notation $\hat a_i$ ($\hat b_i$) for the
site operators of the $A$ ($B$) sites and define
the connectivity number $z_A$ ($z_B$). For the specific
example of the 3$\times$2 cluster, we have
$z_{A}=2$ and $z_{B}=1$. The element $W_{BA}$ is obtained from
$W_{AB}$  by interchanging $A$ and $B$. Since the matrix elements
can be chosen to be real, we in fact have $W_{BA} = W_{AB}$.

The assumption of a two-component order parameter can also be used
at the very beginning of this analysis
to simplify the mean-field perturbation in (\ref{MFP}). One finds
for the example being considered that
\begin{equation}
\label{mf_term_2op}
\hat{\mathcal 
V}^{MF}_{3\times 2}=-z_{A}J\psi_{A}\sum_{i\in{A}}(\hat{a}_{i}^{\dagger}+
\hat{a}_{i})-z_{B} J\psi_{B}\sum_{i\in{B}}(\hat{b}_{i}^{\dagger}+
\hat{b}_{i})+N_{A}z_{A}J\psi_{A}^{2}+N_{B}z_{B}J\psi_{B}^{2}.
\end{equation}
A perturbative treatment of this mean-field perturbation will of
course yield precisely the grand potential in
(\ref{Omega_0_reduced}). However, the direct calculation
of (\ref{Omega_0_reduced}) does not by itself reveal the correct
eigenvalue equation (\ref{W_red}) required in the determination of
the $\undertilde{W}$ eigenvalues.

The eigenvalues of the energy matrix in (\ref{W_red}) are
\begin{equation}
\omega_\pm = \frac{1}{2}(\bar W_{AA}+\bar W_{BB})\pm \sqrt{\left ( 
\frac{\bar W_{AA}-\bar W_{BB}}{2} \right )^2 + \bar W_{AB}^2},
\end{equation}
where $\bar W_{PQ} = W_{PQ}/\sqrt{N_P N_Q}$.
In the limit $J\to 0$, we find that $\omega_+ \simeq z_A J $ and
$\omega_- \simeq z_B J $. Thus, when  symmetry is used to
reduce the order parameter to the two components $\psi_A$ and 
$\psi_B$, we see that the two energy matrix eigenvalues provided
by the reduction increase linearly with $J$ and are positive. In
fact, these two eigenvalues correspond to the two eigenvalues
in Fig.~\ref{fig:3x2psi} which increase positively from zero.
We thus see that
the reduced grand potential has a local minimum for
small $J$. This behaviour should be contrasted with the result
obtained using the full six-component order parameter for which
negative eigenvalues of the energy matrix are found even in
the limit of small $J$ (see Fig.~\ref{fig:3x2psi}). 

To determine the MI-SF phase boundary with
increasing $J$ we must find a zero crossing of one of the
$\omega_\pm$ eigenvalues.
Since $\omega_+ > \omega_-$ for all values of $J$, the
eigenvalue which first goes to zero is $\omega_-$. Thus the location
of the phase boundary is given by the condition
\begin{equation}
\frac{1}{2}(\bar W_{AA}+\bar W_{BB}) - \sqrt{\left (
\frac{\bar W_{AA}-\bar W_{BB}}{2} \right )^2 + \bar W_{AB}^2} = 0,
\end{equation}
which is equivalent to
\begin{equation}
W_{AA} W_{BB} = W_{AB}^2.
\label{phase_boundary}
\end{equation}
We will use this equation in the following section to map out the
MI-SF phase boundary for the two-dimensional square lattice. We
note that the $N_A$ and $N_B$ factors have dropped out in
(\ref{phase_boundary}) as they must since they play no role in
(\ref{W_red}) when $\omega = 0$. In other words, it is sufficient to
look for the zero eigenvalue of $\undertilde{W}^{\rm red}$ in the
determination of the phase boundary. Also,
the ratio of the order parameter components just as one enters the
SF region will be given by
\begin{equation}
\frac{\psi_B}{\psi_A} = \sqrt{\frac{W_{BB}}{W_{AA}}}.
\label{eq:ratio}
\end{equation}

In light of the very similar numerical values of $\psi_A$ and 
$\psi_B$ shown above, it is reasonable to try one final 
simplification, namely, enforcing the identity $\psi_A = \psi_B$. 
When such an order parameter vector is used, the energy and 
stability matrices reduce to scalars and one recovers a single
order-parameter theory. The critical hopping parameter obtained
in this case for $\mu/U=0.4$ is $J_c/U=0.04816$, as compared to 
$J_c/U=0.04815$ obtained by allowing $\psi_A$ and $\psi_B$ to be
different. Clearly, the inequivalence of different boundary
sites in the MSMFT does not in itself adversely affect the 
prediction of the location of the MI-SF phase boundary for a
homogeneous lattice.

\section{Application to homogeneous lattices in 1, 2, and 3 Dimensions}
\label{Homogeneous}

In this section we present our results for bosons on homogeneous 
lattices in $d$ dimensions, focussing on the linear chain, square, and 
simple cubic lattices. We discuss the phase diagrams that are predicted 
by MSMFT, and  the spatial correlations that are incorporated into 
ground-state wave functions found using this approach.

\begin{figure}[!ht] 
\centering{\includegraphics{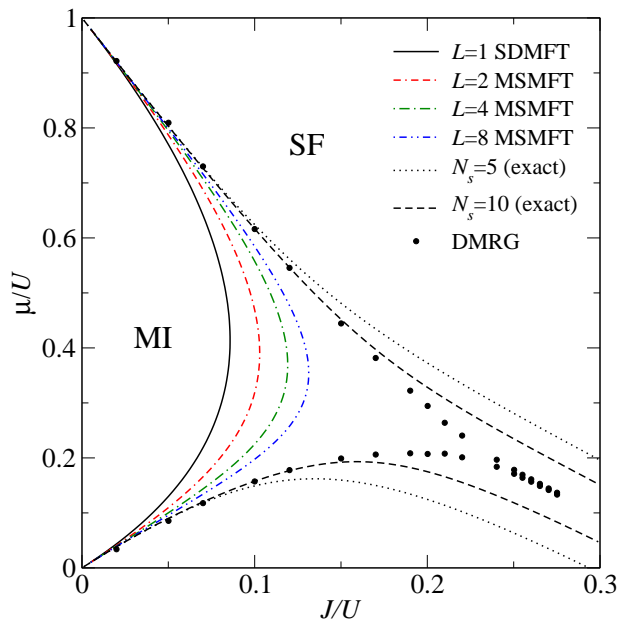}}
  \caption{[Colour online] The MI-SF phase 
  boundaries for a linear chain predicted by MSMFT, compared to 
  the DMRG results of \protect\cite{Kuhner98} shown as circles. 
  The phase boundary systematically approaches the exact result
  with increasing length $L$ of the cluster: $L=2$
  (dash-dash-dot red line),
  $L=4$ (dot-dash green line), $L=8$ (dot-dot-dash blue line). Also shown are the
  boundaries of the $N$-domain where the ground state of
  the full Hamiltonian for a finite chain of length $N_s$ has 
  $N=N_s$ particles.
}\label{fig:1d_phasediag}
\end{figure}

First, we consider the phase diagram of the linear chain in the
region of the first Mott lobe ($0<\mu/U<1$). 
The relevant MSMFT data is 
shown in Fig.~\ref{fig:1d_phasediag}. ``Exact" numerical results, found 
from a DMRG study \cite{Kuhner98}, are also shown for comparison. 
The SDMFT result is seen to be far from the exact phase boundary, 
a result well known in 
the literature~\cite{Sheshadri93,Kuhner98}.
In  Fig.~\ref{fig:1d_phasediag} we also show the results from the application of 
MSMFT for $L$-mers of size
$L=2,4,8$. One can see that as $L$ increases the extent of the Mott 
lobe approaches the DMRG results. Therefore, in one dimension we 
find that there is a systematic improvement in identifying 
the phase boundary upon going from SDMFT to MSMFT. However, the
predicted phase boundaries are still not close to the exact
results and lack the cusp-like feature at the tip of the lobe.
Further improvements could be achieved by increasing $L$ but the
progression in Fig.~\ref{fig:1d_phasediag} suggests that going to $L=16$
would provide only a modest improvement. This value of $L$ is
already beyond our numerical capabilities.

Also shown in Fig.~\ref{fig:1d_phasediag} are results obtained
from an exact diagonalization of the full Hamiltonian in
(\ref{GCBH}) for a finite linear chain with periodic boundary
conditions. Since the Hamiltonian commutes with the total number 
operator, one can determine the $N$-domains in the $\mu/U$-$J/U$
plane where the ground state of $\hat{\mathcal{K}}$
has a total of $N$ particles. For chains of length $N_s$, the
first Mott lobe must be found in the region where 
$N=N_s$, at least in the limit $N_s \to \infty$. 
The boundaries of the $N=N_s$ region are shown in
Fig.~\ref{fig:1d_phasediag} for chains of length $N_s = 5$ and 10. 
It is clear from the figure that the $N=N_s$ domain bounds the
exact Mott lobe with increasing accuracy with increasing $N_s$.
However, exact diagonalizations for much longer chains would be
necessary to accurately represent the tip of the Mott lobe.
 
\begin{figure}[!ht] 
\centerline{\includegraphics{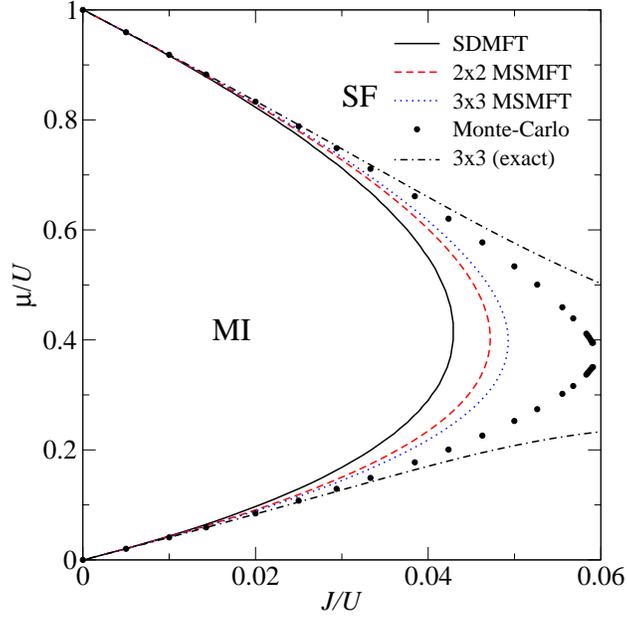}}
\caption{[Colour online] The MI-SF 
phase boundaries for a two-dimensional square lattice, as calculated 
using  SDMFT (solid black line) and MSMFT ($2\times 2$, dashed red 
line; $3\times 3$ (dotted blue line). The MC data (solid black circles) 
are taken from Ref.~\cite{Sansone08}. Also shown (dot-dash black line) 
is the $N$-domain ($N=9$) for the $3\times 3$ cluster determined
by calculating the ground state of the 
full (non-MFT) Hamiltonian with periodic boundary conditions.
}\label{fig:2d_phasediag}
\end{figure}

Our results in two dimensions are shown in Fig.~\ref{fig:2d_phasediag}. 
In the MSMFT calculations we have used clusters of sizes $2\times 2$ 
and $3\times 3$; the ``exact" results
are taken from the Monte Carlo study of Ref. \cite{Sansone08}. 
As in the 
previous figure we also show the boundaries of the $N$-domain
($N=9$ in this case)
for the exact solution of the full Hamiltonian 
of a $3\times 3$ cluster with periodic boundary conditions. To
perform the calculations for the largest cluster it was
necessary to take advantage of the symmetry of the cluster; some
details of the method used are given in the Appendix. Similar 
results are 
shown in Fig.~\ref{fig:3d_phasediag} for three dimensions, where
we have performed calculations for $2\times 2\times 1$ and a 
$2\times 2\times 2$ clusters. 
One sees a progressive improvement of the
phase boundary with increasing cluster size, as in the case of
the linear chain, with the phase boundary approaching the exact 
(MC) results~\cite{Sansone07}. The best agreement is obtained in 
three dimensions
as one might expect on the basis of the validity of mean field
theory when the upper critical dimension is approached.

\begin{figure}[!ht] 
\centerline{\includegraphics{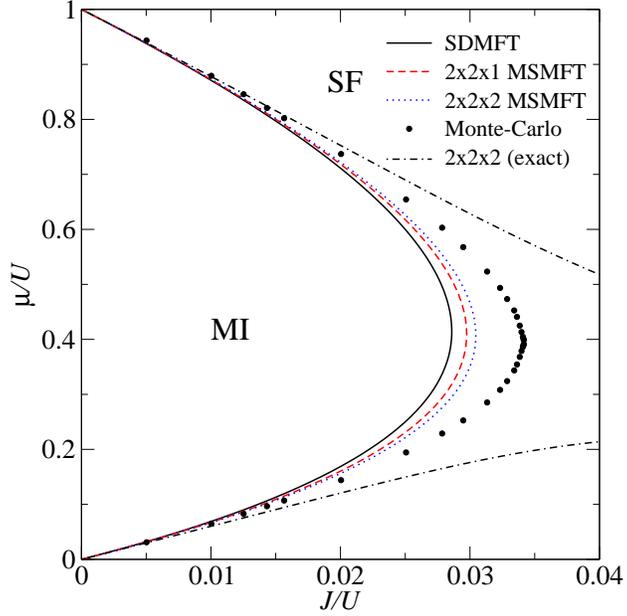}}
  \caption{[Colour online] The MI-SF 
phase boundaries for a three-dimensional cubic lattice, as calculated 
using  SDMFT (solid black line) and MSMFT ($2\times 2\times 1$,
dashed red line; $2\times 2\times 2$, dotted blue line). 
The MC data (solid black circles) taken from Ref.
\cite{Sansone07}.
}
\label{fig:3d_phasediag}
\end{figure}

For the $3\times 3$ cluster in Fig.~\ref{fig:3x3}(b), the $A$ and $B$ 
sites are inequivalent and as a result, the order parameters
$\psi_A$ and $\psi_B$ as one enters the SF phase will be different. 
We find, however, that the ratio given in (\ref{eq:ratio})
deviates from unity by no more than a few percent. It is
therefore of interest to see the effect of enforcing the
equality of $\psi_A$ and $\psi_B$ in the MSMFT calculations.
(Note that this does not reduce MSMFT to SDMFT since one must
still determine the eigenstates of $\hat{\mathcal K}^0_L$ for
the whole cluster.) We find that the phase 
boundaries shown in the two-dimensional phase diagram are
virtually unchanged when this constraint on the order parameters 
is imposed. Thus the calculations for homogeneous lattices can
be simplified by assuming that the order parameters of all the
boundary sites of the cluster are the same.

\begin{figure}[!ht] 
\centerline{\includegraphics{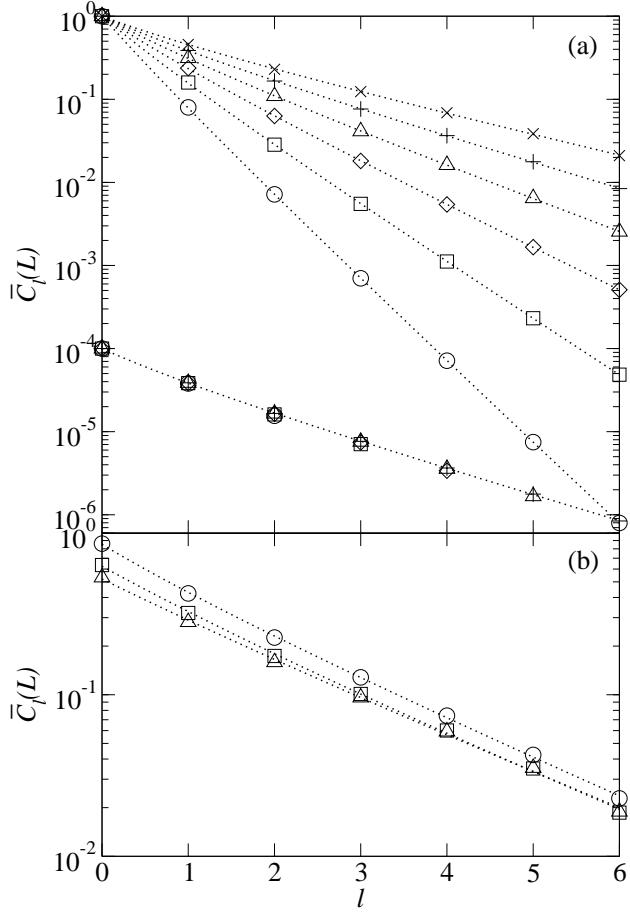}}
  \caption{The average correlation function of the linear chain
within MSMFT versus the intersite spacing. Panel (a) shows
results for $L=7$ in the MI phase: the six
curves starting from 1 at $n=0$ correspond, from bottom to top, 
to $J/U =$ 0.02, 0.04, 0.06, 0.08, 0.10 and 0.12. The points are
the calculated values of $\bar C_l(L)$ and the dashed curves
are plots of the fitting function (\ref{eqn:fits}). The lowest
curve in (a) shows results for $J/U = 0.10$ for $L =$ 3, 4, 5, 6
and 7, all plotted on top of each other. This curve has been
displaced from the others for clarity by multiplying the data by
$10^{-4}$. Panel (b) shows results for $L=7$ in the SF phase:
from top to bottom, $J/U =$ 0.14, 0.17, and 0.20.
}\label{fig:C_n}
\end{figure}

With several sites in the cluster, one can also consider 
\textit{spatial correlations} between the various sites.
As mentioned
in the introduction, this is one of the main advantages of
using a MSMFT.  
The correlations we consider are those exhibited by the
function
\begin{equation}\label{correlations}
C_{ij}=\langle \hat{\phi}^{\dagger}_{i}\hat \phi_{j}\rangle,
\end{equation}
where 
\begin{equation}
\hat\phi_{i}=\hat{a}_{i}-\langle\hat{a}_{i}\rangle.
\end{equation}
In the MI phase, $C_{ij}$ reduces to the single-particle
density matrix while in the SF phase it is
the contribution to the density matrix from the noncondensed
component.

We first consider our results for the linear chain, specifically
along the $\mu/U = 0.4$ line in the phase diagram of
Fig.~\ref{fig:1d_phasediag}. For
a given $L$-mer, we calculate $C_{i,i+l}$ for all possible
values of $i$ and $l$ consistent with the size of the $L$-mer.
In the MI phase we find that $C_{i,i+l}$ has a rather weak
dependence on the position $i$ within the cluster despite the
fact that the states of the cluster are calculated with
open-chain boundary conditions. For this reason we plot only
$\bar C_l(L)$, the average of $C_{i,i+l}$ over $i$. This quantity
is shown in Fig.~\ref{fig:C_n} for $L =7$ for $l$
between 0 and 6 and for values of $J/U$ that span the phase
boundary. Since this log-linear plot reveals a dependence on $l$
which deviates weakly from a pure exponential, we have 
fit this data to the form
\begin{equation}
\bar C_l(L) \sim {e^{-l/\xi}\over (l_0+l)^\eta},
\label{eqn:fits}
\end{equation}
where the fit parameters $l_0$, $\xi$ and $\eta$ are in general
functions of $L$~\cite{footnote1}.
We emphasize
that this form of fitting function is chosen simply because
it describes the {\it short-range} behaviour of the 
correlation function with reasonable accuracy. In the limit
$L\to \infty$, one in principle would be able to determine the
asymptotic behaviour of the correlation function which for large
$l$ should behave as 
\begin{equation}
\bar C_l(\infty) \sim {e^{-l/\xi_\infty}\over l^{\eta_\infty}}.
\end{equation}
At the phase boundary, one expects $\xi_\infty$ to
diverge~\cite{Kuhner98},
resulting in an algebraic decay of the correlation function.
However, the finite size
of the $L$-mers considered in our calculations precludes extracting 
useful information about this
asymptotic behaviour.

We also observe that in the MI phase the values of $\bar C_l(L)$
for a fixed $l$ depend weakly on $L$. This is confirmed by the
lowest curve in Fig.~\ref{fig:C_n}(a)
for $J/U = 0.1$ where we plot all the available data for $\bar
C_l(L)$ for $L = 3,\dots,8$. This shows that the short-range
behaviour of the $\bar C_l(L)$ correlation function is not
affected significantly by increasing the cluster size. The
results in the SF phase are shown for $L=7$ in
Fig.~\ref{fig:C_n}(b). The behaviour shown here is qualitatively
different from the MI phase.

\begin{figure}[!t] 
\centerline{\includegraphics{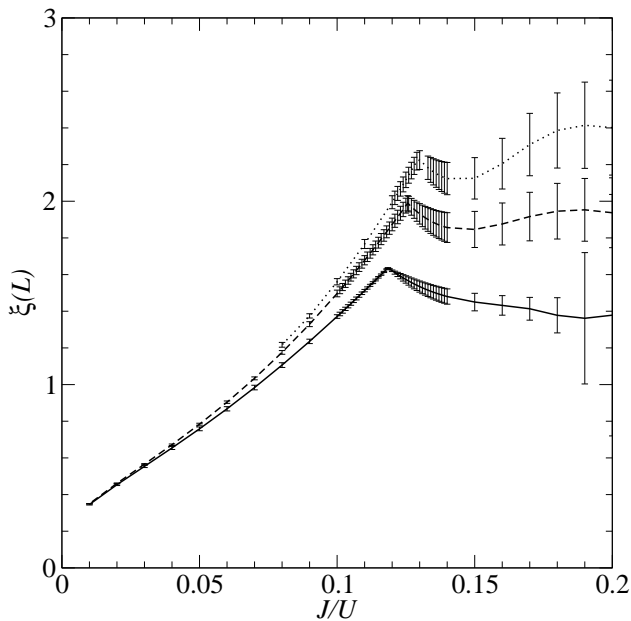}}
  \caption{The correlation length $\xi(L)$ {\it vs.} $J/U$ for
the linear chain of length $L=$
4, 6 and 8 (bottom to top). The results were obtained by fitting 
the data to (\ref{eqn:fits}); the vertical bars indicate the 
estimated error.
}\label{fig:xi}
\end{figure}

In Fig.~\ref{fig:xi} we show the correlation length $\xi(L)$ obtained 
by fits of the data to (\ref{eqn:fits}) as a function of $J/U$.
In the MI phase there is a monotonic increase of $\xi$ with
$J/U$, reaching a peak (or cusp) at the phase boundary between the MI
and SF phases, the position of which of course depends on $L$.
As stated earlier, the variation of $\xi$ with $L$ at a fixed
valued of $J/U$ is relatively weak in the MI phase; part of this 
variation is simply a result of fitting data over an 
increasingly larger range of
$l$. However, the $L$-dependence of $\xi(L)$ is much stronger in
the SF phase. The distinct behaviour of $\xi(L)$ between the MI
and SF phases persists for all values of $\mu/U$. 
As for the other fitting parameters, we find that $l_0
\simeq 1$ in the MI phase, reflecting the fact that $\bar
C_{l=0}(L)\simeq 1$; $l_0$ then increases slightly as $\bar
C_{l=0}(L)$ decreases in the SF phase. Finally, we find that
$\eta(L)$ is close to 0.4 for most of the data considered.

\begin{figure}[!ht] 
\centerline{\includegraphics{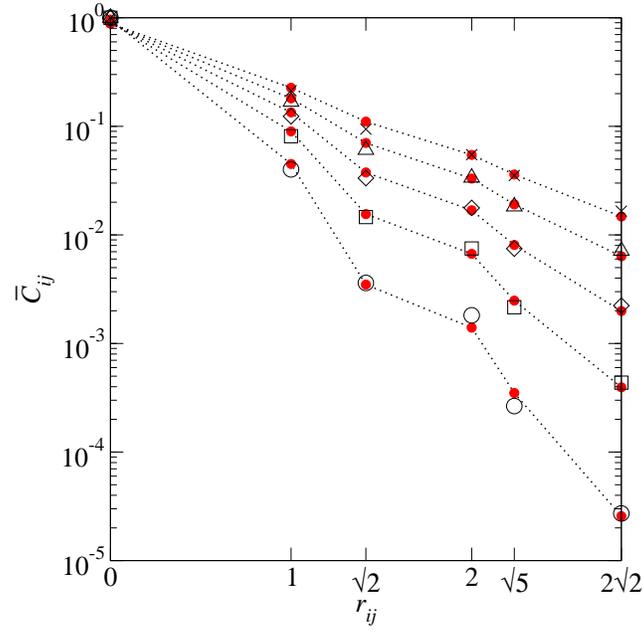}}
\caption{[Colour online] The variation of the averaged pair correlations with 
separation in units of the lattice constant for the
two-dimensional $3\times 3$ cluster. The points
are the result of the MSMFT calculation for
$\mu/U = 0.4$ and for $J/U=0.01$ (circle), 0.02 (square), 
0.03 (diamond), 0.04 (triangle) and 0.049 (cross).
The values of $\bar C_{ij}$ as obtained from the fitting
function in (\ref{eqn:fit_2D}) are also plotted as solid points
(red) which are joined by dotted lines as a
guide to the eye. The MI-SF transition for this
value of $\mu/U$ occurs close to $J/U = 0.0492$.
}

\label{fig:2d_correlationfcn}
\end{figure}

Rather similar behaviour is found for
the two-dimensional $3\times 3$ cluster in 
Fig.~\ref{fig:3x3}(b) in the MI phase.
For this cluster there are five distinct intersite separations
$r_{ij}$, namely $1$, $\sqrt{2}$, $2$, 
$\sqrt{5}$, and $2\sqrt{2}$ in units of the lattice constant.
Due to the finite size of the cluster, there are pairs of sites,
such as the nearest-neighbour $AB$ and $BC$
sites, for which the correlations are not equivalent.
However, the differences are found to be
small and we therefore average the correlations over all pairs of
sites which, by symmetry, would be equivalent in the infinite
lattice. These averaged values are
plotted in Fig.~\ref{fig:2d_correlationfcn}.

\begin{figure}[!ht] 
\centerline{\includegraphics{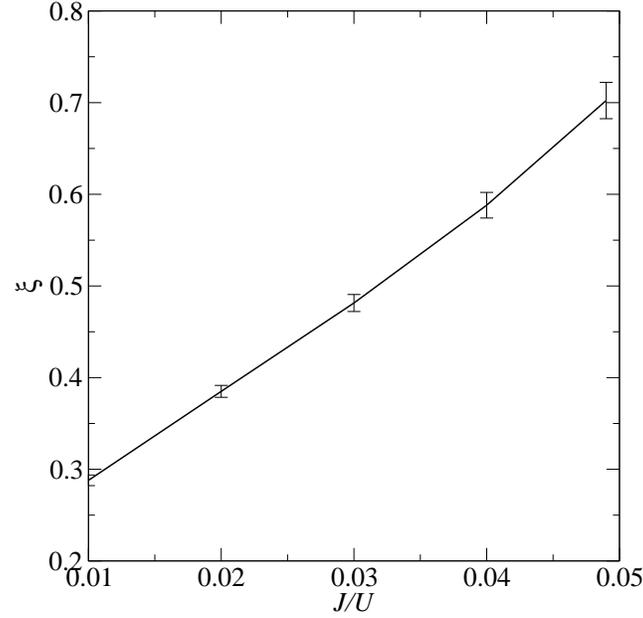}}
  \caption{The dependence of the correlation length, $\xi$ (in
units of the lattice constant), on
$J/U$ for the two-dimensional square lattice using a $3\times 3$
cluster. The results were obtained by fitting the data in
Fig.~\ref{fig:2d_correlationfcn} to (\ref{eqn:fit_2D}).  }
\label{fig:2d_xi}
\end{figure}

The data in Fig.~\ref{fig:2d_correlationfcn} were fit to a
function having the following form
\begin{equation}
\bar C_{ij}=e^{-r_{ij}/\xi}\left (a+b(1-\delta_{ij})\cos4\theta_{ij}\right ),
\label{eqn:fit_2D}
\end{equation}
where $\xi$ is again the correlation length, and the parameters
$a$ and $b$ are introduced to capture the anisotropy of the
lattice correlation function. The variable $\theta_{ij}$ is the
angle that the vector ${\bf r}_{ij}$ makes with the $x$-axis.
The values of $\bar C_{ij}$ as determined by (\ref{eqn:fit_2D})
are also plotted in Fig.~\ref{fig:2d_correlationfcn} for the
optimal values of the fitting parameters. These values are
joined by the dotted lines to provide a guide to the eye. It
can be seen that the data points lie
quite close to these lines for all $J/U$, indicating that the 
assumed form of the angular dependence does a reasonable job of 
representing the data. The correlation length $\xi$ is plotted
in Fig.~\ref{fig:2d_xi} as a function of $J/U$. The increasing
trend is similar to that found for the linear chain, but the
2D correlation length is significantly smaller than in 1D. We
emphasize again that the correlation length we have extracted
represents the short-range behaviour of the correlation
function. One would need the long-range behaviour in order to 
determine the critical
behaviour of the correlation length at the MI-SF phase boundary.

Finally, we consider the occupation number distribution for a
particular site in the $L$-mer in the MI phase. The probability
of finding the configuration $|\{n_l\}\rangle$ in $|0\rangle$, the
ground state of ${\cal K}_L^0$, is $|\langle \{ n_l
\}|0\rangle|^2$. Thus the occupation number distribution for,
say site 1, is 
\begin{equation}
P(n_1) = \sum_{n_2,\cdots,n_l} |\langle \{ n_l\}|0\rangle|^2.
\end{equation}
Near the tip of the first  Mott lobe of the $3\times 3$ $L$-mer of the
2D square lattice ($\mu/U = 0.4$, $J/U = 0.049$) we find $P(0) =
0.022$, $P(1) = 0.955$ and $P(2) = 0.023$ for the central site of the
$3\times 3$ $L$-mer. Even though the average occupancy
of the sites within the $L$-mer is exactly $n=1$ in the MI
phase, the occupancy of a given site does fluctuate about its
average value; the probability of finding 0 or 2 particles on
this site is roughly 2\%.

\section{Application to Superlattices}
\label{Diatomic Results}

\subsection{The Dimer Chain}

\begin{figure}[!ht]
\centering{\includegraphics{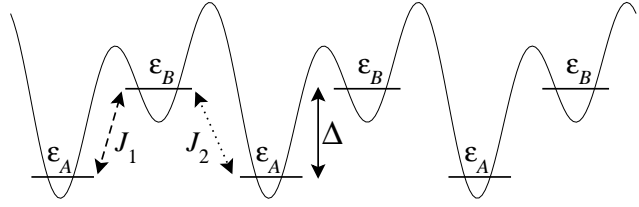}}
\caption{Schematic of a general dimer superlattice with
alternating site energies $\varepsilon_A$ and $\varepsilon_B$.
The level separation is $\Delta$.
The different barrier heights along the chain lead to different
intra-dimer ($J_1$) and inter-dimer ($J_2$) tunnelling energies.
}
\label{fig:dimerchain_gen}
\end{figure}
\begin{figure}[!ht]
\centering{\includegraphics{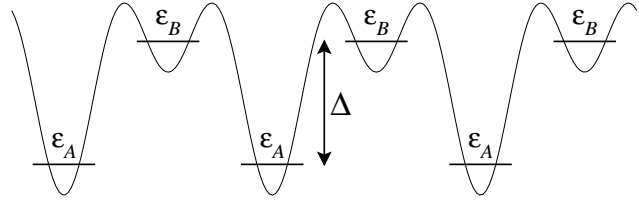}}
\caption{A portion of the dimer superlattice used in the
analysis of the MI-SF transition. The inversion symmetry about
each lattice site results in equal intra- and inter-dimer
tunnelling energies $J$.
}
\label{fig:dimerchain}
\end{figure}

We now turn to an analysis of a simple superlattice
consisting of a 
one-dimensional chain in which the on-site energies alternate in an 
$...\varepsilon_{A}\varepsilon_{B}\varepsilon_{A}\varepsilon_{B}...$ 
fashion. Figure \ref{fig:dimerchain_gen} shows a portion of such
a lattice, with a pair of $AB$ sites being referred to as a 
{\it dimer}. The BH Hamiltonian for this system can then be
expressed as
\begin{eqnarray}\label{Kdia_pre}
&&\hat{\mathcal{K}}=~\frac{\Delta}{2}\sum_{j}\left(\hat{n}_j^B - 
\hat{n}_j^A\right)-~\mu\sum_{j}\left( \hat{n}_j^A + 
\hat{n}_j^B\right)
 +\frac{1}{2} \sum_j 
\left(U_{A}\hat{n}_j^A(\hat{n}_j^A~-~1)+U_{B}
\hat{n}_j^B(\hat{n}_j^B~-~1)\right)\nonumber \\
 &&\hskip 1truein-J_1\sum_{j}\left( \hat{a}^\dagger_{j} 
\hat{b}_{j} + \hat{b}^\dagger_{j} \hat{a}_{j} \right) - J_2 \sum_{ j}
\left( \hat{a}^\dagger_{j+1} \hat{b}_{j} + \hat{b}^\dagger_j
\hat{a}_{j+1} \right)~, 
\end{eqnarray}
where the index $j$ labels the $j$th dimer within the lattice. 
Here, we have introduced the site operators 
$\hat{a}_{j}(\hat{a}^{\dagger}_{j})$ and 
$\hat{b}_{j}(\hat{b}^{\dagger}_{j})$ for the $A$ and $B$ sites, 
respectively, in the $j$th dimer, and the number operators
$\hat{n}_j^A=\hat{a}^{\dagger}_{j}\hat{a}_{j}$ and 
$\hat{n}_j^B=\hat{b}^{\dagger}_{j}\hat{b}_{j}$. 
With $\varepsilon_A = -\Delta/2$ and  $\varepsilon_B =
\Delta/2$, the level separation within a dimer is $\Delta$.
We have also allowed for different interaction parameters $U_A$
and $U_B$ for the two kinds of sites and for different
intra-dimer ($J_1$) and inter-dimer ($J_2$) hopping (tunnelling)
energies. For the optical lattice potential illustrated in
Fig.~\ref{fig:dimerchain_gen}, the asymmetric barrier heights lead
to $J_1 > J_2$.
In the following, we consider the optical lattice shown in 
Fig.~\ref{fig:dimerchain} which has inversion symmetry 
and symmetric barrier heights. In this case, $J_1 = J_2 = J$.
For simplicity, we also assume $U_A = U_B = U$. This particular
model was previously studied in \cite{Rousseau06} and
\cite{Chen10}.

To implement the MSMFT, we partition the lattice into clusters
containing $N_d$ dimers, thus defining $L$-mers of size $L =
2N_d$. Each $L$-mer begins with an $A$ site and ends with a $B$
site, and the different $L$-mers are decoupled by introducing
the order parameters $\psi_A$ and $\psi_B$. Following the 
procedure given in \S~\ref{2D}, we obtain the 
mean-field decoupled $L$-mer Hamiltonian
\begin{equation}\label{KDIA}
\hat{\mathcal K}^{MF}=\hat{\mathcal K}^{0}_{L}+\hat{\cal V}^{MF}_{L},
\end{equation}
where
\begin{eqnarray}\label{K01D_dia}
&&\hat{\mathcal K}^{0}_{L}
=\sum_{j=1}^{N_d}\left[\left(-\frac{\Delta}{2}-\mu\right
)\hat{n}_j^A+\left(\frac{\Delta}{2}-\mu\right)\hat{n}_j^B+\frac{U}{2
}\Big(\hat{n}_j^A\left(\hat{n}_j^A-1\right)+\hat{n}_j^B\left(\hat{
n}_j^B-1\right)\Big)\right]\nonumber\\
&&\hskip 1truein-J\sum_{j=1}^{N_d}(\hat{a}_{j}^{\dagger}\hat{b}_{j}+\hat{b
}_{j}^{\dagger}\hat{a}_{j})-J\sum_{j=1}^{N_d-1}(\hat{a}_{j+1}^{\dagger}
\hat{b}_j+\hat{b}_j^{\dagger}\hat{a}_{j+1})
\end{eqnarray}
is the Hamiltonian of an open-ended dimer chain of length
$L=2N_d$, and
\begin{equation}\label{mf_term_dia}
\hat{\cal V}^{MF}_{L}=-J\psi_{B}\left(\hat{a}_{1}^{\dagger}+\hat{a}_{1}
\right)-J\psi_{A}\left(\hat{b}_{N_{d}}^{\dagger}+\hat{b}_{N_{d}}\right)
+2J\psi_{A}\psi_{B}~
\end{equation}
is the mean-field coupling.

To determine the MI-SF phase boundary we use the energy
criterion based on the eigenvalues of the energy matrix
$\undertilde{W}$. For this example of a two-component order
parameter, the expansion of the grand potential has the form
given in (\ref{Omega_0_reduced}) with the matrix elements
\begin{equation}
W_{AA} = J^2 \sum_{\nu \ne 0}
\frac{|\langle \nu| 
\hat{b}_{N_d}^{\dagger}+\hat{b}_{N_d}|0\rangle|^{2}}
{\Omega_0(\{0\})-\Omega_\nu(\{0\})}~,
\label{W_AA_dimer}
\end{equation}
\begin{equation}
W_{BB} = J^2 \sum_{\nu \ne 0}
\frac{|\langle \nu| 
\hat{a}_{1}^{\dagger}+\hat{a}_{1}|0\rangle|^{2}}{
\Omega_{0}(\{0\})-\Omega_\nu(\{0\})}~,
\label{W_BB_dimer}
\end{equation}
\begin{equation}
W_{AB} = J+ J^2 \sum_{\nu \ne 0} \frac{\langle 0| 
\hat b_{N_d}^\dagger + \hat{b}_{N_d})|\nu\rangle\langle \nu |
\hat a_1^\dagger + \hat{a}_{1}|0\rangle}
{ \Omega_{0}(\{0\})-\Omega_\nu(\{0\})}~,
\end{equation}
and $W_{BA} = W_{AB}$. The $\undertilde{W}$ eigenvalues in this
case are given by
\begin{equation}
\omega_\pm = \frac{1}{2}(W_{AA}+W_{BB})\pm \sqrt{\left ( 
\frac{W_{AA}- W_{BB}}{2} \right )^2 + W_{AB}^2}.
\end{equation}
Since $W_{AA}$ and $W_{BB}$ are negative definite, we see that
$\omega_- < 0$. On the other hand,
since $W_{AB} \simeq J$ for small $J$, while $W_{AA}$
and $W_{BB}$ are proportional to $J^2$, we see that $\omega_+$
remains positive with increasing $J$ up to the point where 
\begin{equation}
W_{AA}W_{BB} = W_{AB}^2,
\end{equation}
when it goes to zero. 
This equation defines the phase boundary between the MI and the
SF phases. We thus see that 
$\boldsymbol{\psi}={\bf 0}$ is a saddle
point of the grand potential in the MI phase and that this point 
turns into a local maximum when the SF phase is entered. The
stationary point remains a saddle point in the SF phase.

\subsubsection{Results for $N_d=1$}

\begin{figure}[!ht]
\centering{\includegraphics[scale=0.55]{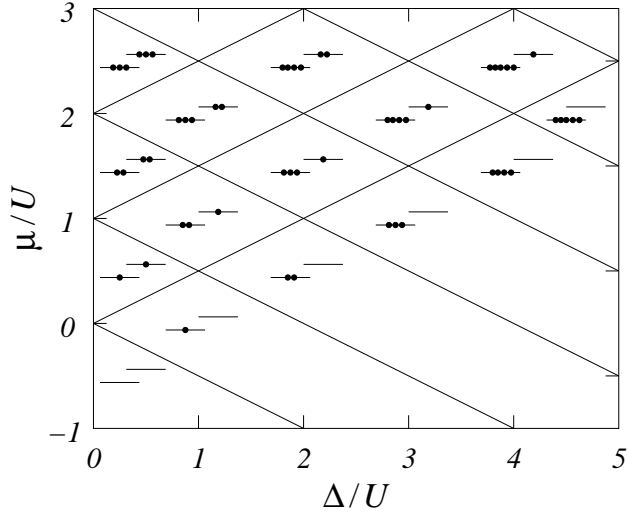}}
\caption{ The $N$-domains of a single-dimer $L$-mer at 
$\tilde{J}=0$. The site occupations of the $A$ (lower level) and
$B$ (upper level) sites is indicated by the number of black dots.
As one crosses a downward (upward) sloping line with increasing
$\mu/U$, the occupation of the lower (upper) level increases by
one. This figure applies to any number of dimers in the $L$-mer
being considered.}
\label{fig:1dimerJD0}
\end{figure}

In this subsection we present results for the case of an $L$-mer
consisting of a single dimer, that is, $N_d=1$. This
case already exhibits all the general features of the phase
diagram for a dimer chain. The results we obtain in the
following subsection show quantitative
improvements with increasing $N_d$ but are qualitatively similar.  
The results we find for $N_d=1$ are also similar to those found
in an earlier investigation~\cite{Chen10}. This latter work, 
however, is a SDMFT (even though different order 
parameters are introduced for the $A$ and $B$ sites) since
the dimer Hamiltonian used in the perturbative analysis is
the sum of independent site Hamiltonians. In contrast, ours is
truly a MSMFT, even for $N_d=1$, since it
requires the determination of the eigenstates of the coupled
dimer Hamiltonian in (\ref{K01D_dia}).

The phase diagram for the dimer chain is determined
in the three-dimensional parameter space defined by
the dimensionless variables $\tilde \mu = \mu/U$,
$\tilde J = J/U$ and $\tilde \Delta = \Delta/U$. For given
values of $\tilde \mu$ and $\tilde \Delta$, the system is in a
MI phase for $\tilde J < \tilde J_{\rm cr}(\tilde \mu, \tilde
\Delta)$. This function defines the phase boundary as a surface
in the three-dimensional parameter space. As we shall see, this 
surface consists of sections, each of which terminates on the
$\tilde J = 0$ plane and corresponds to distinct MI regions.
In order to understand the underlying structure of the phase
diagram, it is therefore useful to first consider the limiting case 
of $\tilde{J}=0$ which defines the base of the Mott lobes.
For $\tilde{J}=0$ and $N_d
=1$, we have the dimensionless dimer Hamiltonian
(we can dispense with the dimer index $j$ in (\ref{K01D_dia})
since there is a single dimer in the $L$-mer)
\begin{equation}
\tilde{\mathcal K}^{0}_{L}(\tilde J=0)
=\left(-\frac{\tilde{\Delta}}{2}-\tilde \mu\right)\hat{n}_A
+\left(\frac{\tilde{\Delta}}{2}-\tilde \mu\right)\hat{n}_B
+\frac{1}{2}
\left(\hat{n}_A\left(\hat{n}_A-1\right)+\hat{n}_B
\left(\hat{n}_B-1\right)\right).
\end{equation}
The occupation number states $|n_A,n_B\rangle$ are eigenstates
of this Hamiltonian with eigenvalues
\begin{equation}\label{K0J0}
\tilde \kappa_L^0 =\frac{1}{2}\left(n_{A}-\tilde{\mu}_{A}
-\frac{1}{2}\right)^{2}-\frac{1}{2}\left(\tilde{\mu}_{A}+\frac{1}{2}
\right)^{2}+\frac{1}{2}\left(n_{B}-\tilde{\mu}_{B}
-\frac{1}{2}\right)^{2}
-\frac{1}{2}\left(\tilde{\mu}_{B}+\frac{1}{2}\right)^{2},
\end{equation}
where $\tilde \mu_A = \tilde \mu + \tilde \Delta/2$ and $\tilde
\mu_B = \tilde \mu - \tilde \Delta/2$. The lowest energy is
obtained by minimizing $\tilde \kappa_L^0$ with respect to $n_A$
and $n_B$. Since $n_A$ is an integer, the minimizing value of
$n_A$ is the integer closest to $\tilde \mu_A +1/2$, that is,
the integer $\nu_A$ when $\nu_A -1 < \tilde
\mu_A < \nu_A$. This condition defines a region in the
$\tilde\mu$-$\tilde \Delta$ plane where the number of atoms on
the $A$ site is $\nu_A$. These regions are shown in
Fig.~\ref{fig:1dimerJD0} and
are bounded by the lines $\tilde \mu = \nu_A - \tilde \Delta/2$
for $\nu_A = 0,\,1,\cdots$. A similar analysis determines $\nu_B$,
the minimizing value of $n_B$, and the regions where this value
applies are bounded by the lines $\tilde \mu = \nu_B + \tilde
\Delta/2$ with $\nu_B = 0,\,1,\cdots$. As shown in
Fig.~\ref{fig:1dimerJD0}, the
$\tilde \mu$-$\tilde \Delta$ plane is thus divided into domains
in which the number of particles in the dimer is $N =
\nu_A+\nu_B$. These domains are the $N$-domains introduced
earlier and define the base of the distinct MI regions. We observe
that each of these regions has specific site occupations. For
example, for $N = 2$, there is a region where the dimer
configuration is given by $\nu_A = 1,\,\nu_B=1$ and another
with $\nu_A = 2,\,\nu_B=0$. These two configurations are
degenerate in energy at a single point in the $\tilde
\mu$-$\tilde \Delta$ plane. They are also degenerate with other
configurations having a different value of $N$ along the lines
bounding the $N$-domains. With increasing $\tilde J$, the
occupation number states are no longer eigenstates of the dimer
Hamiltonian but within each Mott lobe the dimer 
configuration is predominantly that indicated
in Fig.~\ref{fig:1dimerJD0} at the base of the lobe.

\begin{figure}[!ht]
\centering{\includegraphics{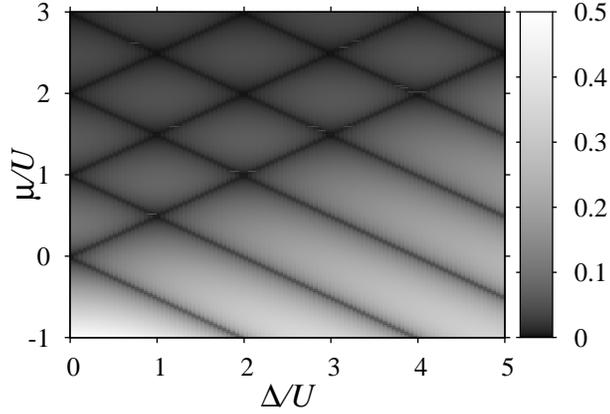}}
\caption{The phase diagram of a dimer chain
within the MSMFT using a cluster consisting of a 
single dimer. The grey scale indicates the value of $\tilde
J_{\rm cr}(\tilde \mu, \tilde \Delta)$ at which the MI-SF phase
boundary occurs. Each region bounded by a black contour ($\tilde
J = 0$) corresponds to a different Mott lobe with an integer
number of atoms $N$ in the dimer. }
\label{fig:1dimercontour}
\end{figure}

In Fig.~\ref{fig:1dimercontour} we show a 
contour plot of the $\tilde J_{\rm cr}(\tilde \mu,\tilde \Delta)$ 
surface, below which the MI phase exists. The grey scale
indicates the extent of the MI phase in the $\tilde J$ direction
which is perpendicular to the plane of the figure. In general,
we see that the area enclosed by a given contour decreases as
$\tilde J$ increases, resulting in a dome-like structure of the
MI-SF phase boundary. The $\tilde J = 0$ contour (black)
corresponds to the lines in Fig.~\ref{fig:1dimerJD0} 
and defines the base of each
of the Mott lobes. Within each lobe the number of particles
in the dimer, $\langle \hat n_A\rangle + \langle \hat
n_B\rangle$, is equal to the integer $N$ indicated by the
configuration in Fig.~\ref{fig:1dimerJD0}. 
However, as we show in more detail
later, the average site occupancies $\langle \hat n_A\rangle$
and $\langle \hat n_B\rangle$ are not constant throughout a
given lobe.
 
\begin{figure}[!ht]
\centering{\includegraphics{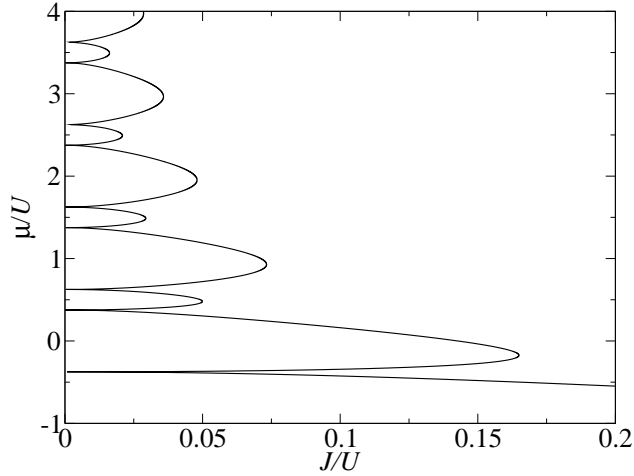}}
\caption{A slice through the phase diagram in
Fig.~\ref{fig:1dimercontour} at $\tilde \Delta = 0.75$.
The average filling per site for each Mott lobe increases by
$1/2$ with increasing $\mu/U$, starting with $n = 0$ at the
bottom of the figure. The system is a SF to the right of the
phase boundary.
}
\label{fig:tilde_Delta_slice}
\end{figure}

Although Fig.~\ref{fig:1dimercontour} provides the overall
structure of the phase diagram, it is useful to consider various
two-dimensional cross-sections to visualize the MI and SF
regions. A slice through Fig.~\ref{fig:1dimercontour} at a
constant value of $\tilde J$ gives one of the contours in the
figure and reveals the Mott regions as islands in the $\tilde
\mu$-$\tilde \Delta$ plane surrounded by SF regions. Another
possibility is a vertical slice through Fig.~\ref{fig:1dimercontour}
at a fixed value of $\tilde \Delta$. In
Fig.~\ref{fig:tilde_Delta_slice} we show such a slice at
$\tilde \Delta =0.75$. Together with Fig.~\ref{fig:1dimercontour}
it is easy to visualize the variation of the MI regions with
variations in $\tilde \Delta$.

Perhaps a more physical situation is that for a fixed optical
lattice, hence fixed values of $J$ and $\Delta$, but with
varying $U$ which can be controlled experimentally by means of a
Feshbach resonance. A cross-section at fixed $J/\Delta$ (i.e.,
$\tilde J/\tilde \Delta$) 
corresponds to a plane containing the $\tilde \mu$ axis and
inclined at some angle with respect to the $\tilde J$ and
$\tilde \Delta$ axes. In Fig.~\ref{fig:1dimerJD005} 
we show the intersection of such
a plane with the Mott lobes in Fig.~\ref{fig:1dimercontour} for
$J/\Delta=0.05$. The superfluid phase is indicated by the `shaded' 
region (the vertical black lines) surrounding the various MI
phases that occur for different values of $N$. One can see
that odd values of $N$, corresponding to a half-integral
average number of particles per site, give rise to lobes which
emerge at integral values of $\tilde \mu$. These lobes should
not be confused with the so-called ``loopholes" discussed
in~\cite{Buonsante04a,Buonsante05a} which arise for a different 
reason when more complex superlattice structures are considered.

Also shown in Fig.~\ref{fig:1dimerJD005} are the various
$N$-domains which are bounded by the solid curves. These
domains evolve continuously from those shown in
Fig.~\ref{fig:1dimerJD0} as the $J/\Delta = 0$ plane is tilted
into the $J/\Delta = 0.5$ plane. As one can see, the crossings
of the domain boundaries in Fig.~\ref{fig:1dimerJD0} become
avoided crossings as a result of the finite value of $J$. These
avoided crossings become more pronounced with increasing $\tilde
\Delta$ since $\tilde J = 0.05 \tilde \Delta$.

\subsubsection{Results for $N_d>1$}

\begin{figure}[!ht]
\centering{\includegraphics{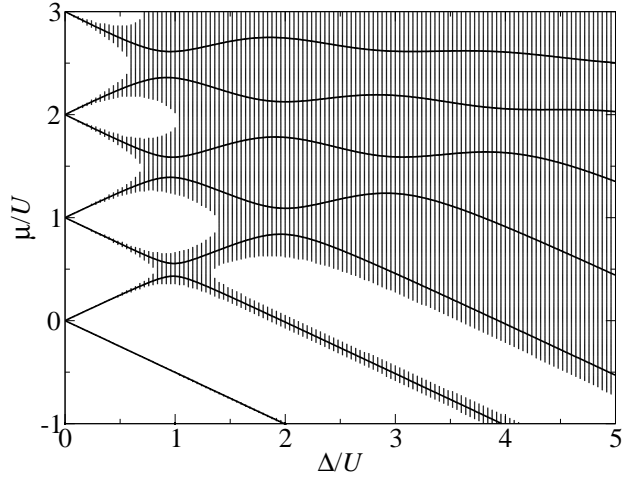}}
\caption{The phase diagram of the dimer chain obtained for
an $L$-mer consisting of a single dimer at $J/\Delta=0.05$. The 
SF phase is indicated by the shaded region (the vertical 
black lines) which surround the various MI regions. 
The black lines are the boundaries of the $N$-domains within
which Mott regions with $N$ particles per dimer are located.
As discussed in the text, the $N$-domains are the regions
where the ground state of 
$\hat{\mathcal K}^{0}_{L}$ contains $N$ particles.}
\label{fig:1dimerJD005} 
\end{figure}

\begin{figure}[!ht]
\centering{\includegraphics{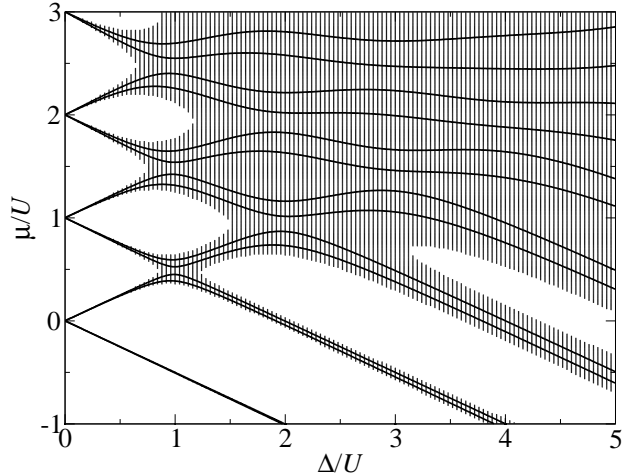}}
\caption{The phase diagram of the dimer chain at
$J/\Delta=0.05$ for an $L$-mer containing two dimers. 
As compared to the single dimer phase diagram at 
in Fig. \ref{fig:1dimerJD005}, we see that the Mott regions are
larger, implying the increased stabilization of the Mott phase.
This is particularly evident in the bottom-right corner of the
figure which shows an additional Mott region with filling
$n=3/2$.
}

\label{fig:2dimers}
\end{figure}

We next consider the effect of increasing the number of dimers
in the $L$-mer to $N_d =2$. The $\tilde J = 0$ phase diagram
is the same as in
Fig.~\ref{fig:1dimerJD0} but the site occupancies shown should be
understood to apply to both dimers so that the number of
particles in each of the $N$-domains is actually doubled.
The difference from $N_d=1$ becomes apparent for $\tilde J \ne
0$ and in Fig.~\ref{fig:2dimers} we show the phase diagram
for $J/\Delta = 0.05$. This phase diagram is very similar to the
$N_d=1$ phase diagram in Fig.~\ref{fig:1dimerJD005} apart
from the appearance of a new Mott region for $N =6$
corresponding to an average
filling ($n = N/L$) of $n=3/2$ particles per site. This simply
reflects the fact that the Mott regions are stabilized with
increasing $L$-mer size as found for the homogeneous systems
discussed in \S \ref{Homogeneous}. As $N_d$ increases, $\tilde
J_{\rm cr}(\tilde \mu, \tilde \Delta)$ increases and the Mott
lobes in Fig.~\ref{fig:1dimercontour} extend to larger values of
$\tilde J$. The intersection of the $J/\Delta = 0.05$ plane with
the Mott lobes can thus include additional Mott regions as found
in the present example.

The other qualitative difference in Fig.~\ref{fig:2dimers}
concerns the $N$-domains which are doubled in number as
compared to those shown in Fig.~\ref{fig:1dimerJD005}. Depending
on the value of $\mu$ the ground state of the Hamiltonian in
(\ref{K01D_dia}) can have $N = 0,1,2,\cdots$ particles and
each of these values appears as a distinct $N$-domain in
Fig.~\ref{fig:2dimers}. Since $L=2N_d =4$, the average
number of particles per site in the $L$-mer is $n = N/4$. 
The additional $N$-domains, however, have no qualitative
effect on the phase diagram and Mott phases only appear at 
integral and half-integral values of $n$. We have performed some
additional calculations for $N_d =3$, that is for $L$-mers with
three dimers. As expected, we find 
only slightly expanded MI regions resulting from the increased 
stabilization of the Mott phases.

\begin{figure}[!ht]
\centering{\includegraphics{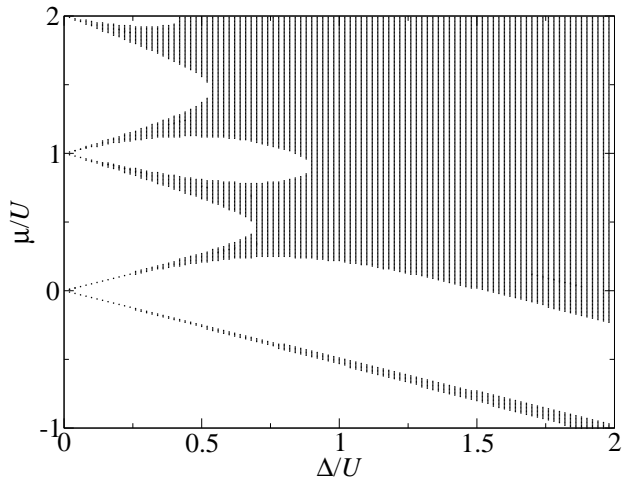}}
\caption{The phase diagram of the dimer chain at
$J/\Delta=0.1$ for an $L$-mer containing two dimers. The range
of $\mu/U$ corresponds to that in Fig.~17 of
Ref.~\cite{Rousseau06} (note that their interaction parameter is
$U' = U/2$). The lowest unshaded region in the figure is a $n=0$
Mott insulator.
}

\label{fig:2dimers_0.1}
\end{figure}

It is of interest to compare the MSMFT phase diagram with
that given in Fig.~17 of Ref.~\cite{Rousseau06} which
was obtained using the quantum Monte Carlo (QMC)
method. These results can effectively be taken as exact. In
terms of our variables, the data in this figure correspond to
$J/\Delta = 0.25$, a factor of five larger than the value used
in Figs.~\ref{fig:1dimerJD005} and \ref{fig:2dimers}. Before
comparing with the data in Ref.~\cite{Rousseau06},
we first note that the extent of the MI regions in the QMC simulations
for 1D systems are enhanced significantly as compared to the
MSMFT results (as seen, for example, in
Fig.~\ref{fig:1d_phasediag}). In other words,
the $\tilde J_{\rm cr}(\tilde\mu,\tilde\Delta)$ surface of the QMC
calculations lies considerably above that of the MSMFT calculations.
As a result, a direct comparison of the results obtained using
the two methods is not particularly meaningful for one and the 
same fixed value of
$J/\Delta$. However, if $J/\Delta$ is scaled in proportion to the
relative extent of the MI regions, one might expect a
correspondence between the two sets of results. This is indeed
the case. The results in Fig.~17 of Ref.~\cite{Rousseau06}
are qualitatively similar to those of Fig.~\ref{fig:2dimers_0.1}
obtained for $J/\Delta = 0.1$. One obvious qualitative
difference is that the MI regions in the MSMFT calculations do
not exhibit the pointed shape seen in the QMC results as was
found for the case of the homogeneous 1D lattice (see
Fig.~\ref{fig:1d_phasediag}).
One can infer from this feature that the QMC version of 
Fig.~\ref{fig:1dimercontour}
would have cusps running along the top of the Mott lobes.
Such cusps are absent in higher dimensions.

Finally, it is instructive to consider the variation of the order
parameters $\psi_A$ and $\psi_B$ and the on-site densities
$\langle \hat n_A\rangle$ and $\langle \hat n_B\rangle$ as a
function of $\tilde \Delta$ along the line $\tilde \mu = 0.5$ in
Fig.~\ref{fig:2dimers}. The order parameters are determined
self-consistently using the iterative scheme in
(\ref{eq:iterative_map}). In the MI
phase, the order parameters converge to zero and the Hamiltonian
$\hat{\mathcal K}_L$ reduces to the dimer Hamiltonian 
$\hat{\mathcal K}_L^0$. Thus in the Mott regions, $\langle \hat n_A
\rangle$ is equal to $\langle \hat n_A \rangle_0$, the value 
obtained for the ground state of $\hat{\mathcal K}_L^0$, with a 
similar equality for the $B$
site. In the SF regions, $\langle \hat n_A \rangle$ is no longer
equal to $\langle \hat n_A \rangle_0$ but the latter quantity is
still well-defined and it is of interest to compare it with the
actual on-site density $\langle \hat n_A\rangle$.

\begin{figure}[!ht]
\centering{\includegraphics{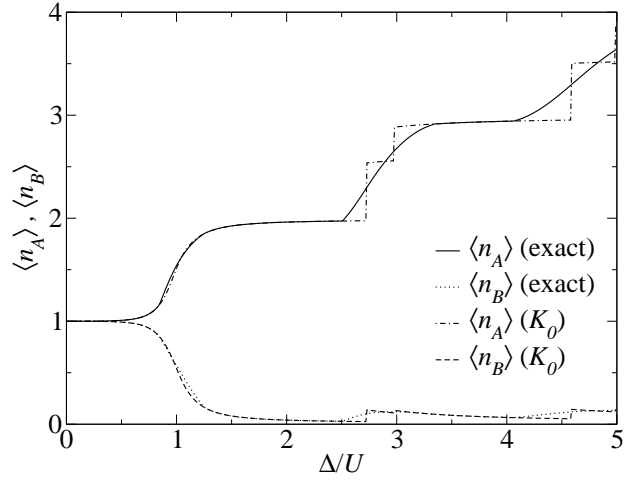}}
\caption{
A plot of the on-site densities for the terminating $A$ and $B$
sites of an $L$-mer containing two dimers, as a 
function of $\tilde{\Delta}$, for the values of $J/\Delta=0.05$ and 
$\tilde{\mu}=0.5$. The densities 
$\langle \hat n_{A}\rangle$  and $\langle \hat n_{B}\rangle$ 
are obtained from the ground state of the full 
Hamiltonian (\ref{KDIA}); for comparison, we also show the
densities
$\langle \hat n_{A}\rangle_0$  and $\langle \hat n_{B}\rangle_0$ 
corresponding to the ground state of $\hat{\mathcal 
K}^{0}_L$ in (\ref{K01D_dia}). 
The discontinuities in $\langle \hat n_{A}\rangle_0$  and
$\langle \hat n_{B}\rangle_0$ occur at the boundaries of the
$N$-domains. On the other hand, $\langle \hat n_{A}\rangle$  and
$\langle \hat n_{B}\rangle$ vary continuously, but show kinks at
the boundaries between the MI and SF phases.
}

\label{fig:A_B_density} 
\end{figure}

In Fig.~\ref{fig:A_B_density} we plot the on-site densities 
for the $A$ and $B$ sites terminating the $L$-mer, as a function of 
$\tilde{\Delta}$, for 
$J/\Delta=0.05$ and $\tilde{\mu}=0.5$. We first consider
$\langle \hat n_A \rangle_0$ and $\langle \hat n_B \rangle_0$ which are
shown by the dashed curves in Fig.~\ref{fig:A_B_density}.
Referring
to Fig. \ref{fig:2dimers}, we see that the $\tilde \mu = 0.5$
line remains in the $N =4$ domain ($n= 1$) for $\tilde \Delta$
between 0 and approximately 3. For this range of $\tilde \Delta$, 
$\langle \hat n_A \rangle_0$ and $\langle \hat n_B \rangle_0$ vary
continuously with $\tilde \Delta$.
For small $\tilde \Delta$, they are both
close to 1 but as $\tilde\Delta$ passes through 1,  
$\langle \hat n_{A}\rangle_0$ increases continuously to about 2 while
$\langle \hat n_{B}\rangle_0$ decreases to about 0. This change
in occupation reflects the change in the dimer state from
predominantly $|1,1\rangle$ to $|2,0\rangle$ as
Fig.~\ref{fig:1dimerJD0} for 
$J/\Delta =0$ would suggest. For larger $\tilde \Delta$, the 
$\tilde \mu=0.5$ line successively enters $N=5$, 6 and 7
domains.  When an $N$-domain boundary is crossed, the densities
$\langle \hat n_A \rangle_0$ and $\langle \hat n_B\rangle_0$
calculated for the ground state of $\hat{\mathcal K}_L^0$ change
discontinuously. On entering the $N=5$ domain, $\langle \hat n_A
\rangle_0$ jumps approximately to 2.25 (the occupancy of
the interior $A$-site jumps to about 2.75).
When the $N=6$ domain is entered, $\langle \hat n_A \rangle_0$
jumps to approximately 3. The latter
indicates that the $\hat {\mathcal K}_L^0$ ground state now
has predominantly the $|3,0\rangle$ configuration.

The full curves in Fig.~\ref{fig:A_B_density} show the variation
of $\langle \hat n_A\rangle$ and $\langle \hat n_B\rangle$.
These quantities generally follow the variation of
$\langle \hat n_A\rangle_0$ and $\langle \hat n_B\rangle_0$, 
respectively, and, as stated earlier, are equal to the latter
in the MI phases where $\langle \psi_A \rangle$ and $\langle
\psi_B\rangle$ are zero. Unlike $\langle \hat n_A\rangle_0$ and
$\langle \hat n_B\rangle_0$, however, $\langle \hat n_A\rangle$
and $\langle \hat n_B\rangle$ vary continuously through the SF regions
from one MI phase to another since in these regions the ground 
state of $\hat {\cal K}^{MF}$ is not a number eigenstate.
We also find that the densities at the interior $A$ and $B$
sites do not deviate much from the densities at the terminating
$A$ and $B$ sites of the $N_d =2$ $L$-mer.

\begin{figure}[!ht]
\centering{\includegraphics{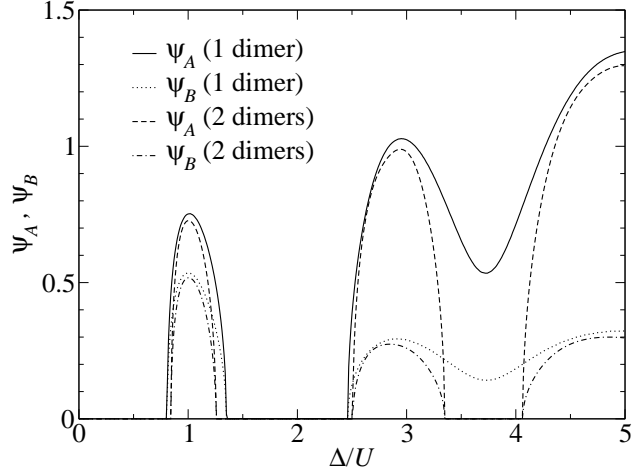}}
\caption{A plot of the order parameters, $\psi_{A}$ and $\psi_{B}$, as 
a function of $\tilde{\Delta}$ for $J/\Delta=0.05$ 
and $\tilde{\mu}=0.5$. The order parameters are shown for
$L$-mers containing one and two dimers corresponding to the phase
diagrams in Figs.~\ref{fig:1dimerJD005} and \ref{fig:2dimers}.
}

\label{fig:2dimerJD005orderparams} 
\end{figure} 

In Fig.~\ref{fig:2dimerJD005orderparams} we plot the order parameters, 
$\psi_{A}$ and $\psi_{B}$, as a function of $\tilde{\Delta}$ for the 
same values of $J/\Delta=0.05$ and $\tilde{\mu}=0.5$. For
comparison, we show the order parameters for both the $N_d = 1$
and $N_d = 2$ $L$-mers corresponding to
Figs.~\ref{fig:1dimerJD005} and \ref{fig:2dimers}. The $\tilde \mu
= 0.5$ line passes through two MI regions in the case of $N_d = 1$
where the order parameters are zero, while in the case of
$N_d=2$, the line passes through three MI regions. The proximity
of a MI region in the $N_d=1$ case leads to the dip in the order
parameters seen in the range $\tilde \Delta \simeq$ 3.5-4. With
increasing $N_d$, the MI phase is stabilized and results in the
$\tilde \mu = 0.5$ scan passing through a MI region for this range of
$\tilde \Delta$.

In the vicinity of a MI-SF boundary the order parameters are seen
to behave as $|\tilde \Delta - \tilde \Delta_{\rm cr}|^{1/2}$
which is consistent with the mean-field behaviour given by
(\ref{eq:psi_mf}). We also observe that the value of $\psi_{A}$ 
is larger than that of $\psi_{B}$, implying a larger superfluid 
fraction on the $A$-site as compared to the $B$-site. This is to
be expected given the lower on-site energy $\varepsilon_A$.
The condensate density ratio $(\psi_B/\psi_A)^2$, however, is
somewhat larger than what one might expect on the basis of the
on-site densities. For example, at $\tilde \Delta = 1$ we have
$(\psi_B/\psi_A)^2 \simeq 0.44$ while $\langle \hat n_B \rangle
/\langle \hat n_A \rangle \simeq 0.33$. A similar enhancement of
the condensate ratio is observed near $\tilde \Delta =3$ and
$\tilde \Delta =5$. 

\newpage

The above results were found for a one-dimensional system, and 
while one does not expect mean-field theory to represent 
the exact ground state in this case, it is simplest to
demonstrate the application of MSMFT to lattices with a 
non-monatomic basis in one dimension.
We wish to stress, however, that the qualitative behaviour obtained for
the dimer chain is also obtained for higher-dimensional
lattices. As one example, we have considered the two-dimensional 
honeycomb lattice in which the site energies within
a single hexagon alternate $ABABAB$. The application of MSMFT 
to this system leads to results which are essentially identical
to those obtained for the dimer chain. For 
a value of $J/\Delta=0.05$ we find a phase diagram similar to 
that shown in Fig.~\ref{fig:2dimers_0.1};
the one minor difference is the extent of the various Mott lobes. 
As for the dimer chain problem,
the phase diagram for the two-dimensional lattice can be understood 
in terms of the $N$-domain structure shown in Fig.~\ref{fig:1dimerJD0}.

\subsection{A Superlattice with Loopholes}

In all of the examples we have considered so far, the phase
diagrams obtained using SDMFT and MSMFT are qualitatively similar. 
As shown previously by Buonsante {\it et
al.}~\cite{Buonsante04a,Buonsante05a}, this is not
always the case. In particular, they find that certain superlattices
exhibit a so-called `loophole' structure which corresponds to Mott
domains that do not arise in SDMFT.

\begin{figure}[!ht]
\centering{\includegraphics{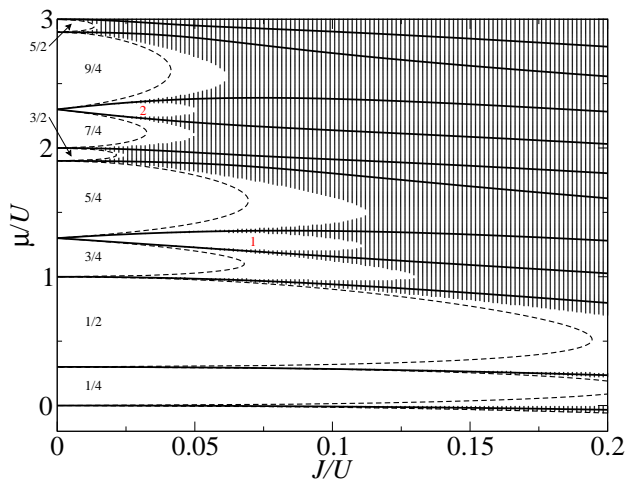}}
\caption{Phase diagram for a four-site superlattice which
exhibits loopholes. The parameters defining the superlattice
are given in the text. The shaded region corresponds to the SF
phase as determined using MSMFT. The dashed lines show the phase
boundaries determined using SDMFT. The numbers $n$ labelling
each of the Mott domains is the average filling of the
superlattice. We note that new Mott domains appear at $n=1$ and
$n=2$ within MSMFT.
The solid lines show the $N$-domains for the four-site
cluster. Within SDMFT, the $N$-domains boundaries are straight
lines, parallel to the $J/U$ axis, which start at the edges of the
Mott domains along the $\mu/U$ axis.}

\label{fig:loopholes} 
\end{figure}

Here we briefly consider an example of this kind. The specific
superlattice we study was previously examined using
SDMFT~\cite{Buonsante04b}. It consists of a four-site superlattice
with site energies given by
$\epsilon_1 = 1.9\Delta$, $\epsilon_2 = 0.3\Delta$, $\epsilon_3
= 1.3\Delta$, $\epsilon_4 = 0.0\Delta$ with the same hopping
parameter $J$ between nearest-neighbour sites and the same
on-site interaction $U$. The phase diagram as determined in
SDMFT and MSMFT is shown in Fig.~\ref{fig:loopholes}. We observe
that Mott domains for integral average filling do not arise with
SDMFT. However, within MSMFT, these domains are present and have
the loophole structure referred to above.

The origin of the integral-$n$ loopholes is associated with
degeneracies of different particle configurations in the atomic
limit ($J\to 0$). Denoting the $\hat {\cal K}^0_L(J=0)$
eigenstates as $|n_1,n_2,n_3,n_4\rangle$, the $N=4$ configurations 
$|0,2,0,2\rangle$ and $|0,1,1,2\rangle$ have the same energy of
$E_4 = 2.6U$. As the chemical potential $\mu$ is swept through
$E_4-E_3 = 1.3U$, the particle number in the system jumps by 2.
This explains the absence of a $n=1$ Mott domain along the
$J/U=0$ axis. As $J/U$ increases, the degeneracy between the two
$N=4$ configurations is lifted within MSMFT and a $n=1$ region
becomes accessible for the formation of a Mott domain. Such a
Mott domain does indeed form as seen in
Fig.~\ref{fig:loopholes}. Within SDMFT, the degeneracy persists
since $\hat {\cal K}^0_L(J=0)$ is used in the perturbation
analysis for all $J$. As a result, the $n=1$ Mott domain does
not appear. A similar argument applies to the $n=2$ Mott domain.
Underlying its formation is the degeneracy of the $|1,3,1,3\rangle$ 
and $|1,2,2,3\rangle$ $N=8$ states in the atomic limit. 

More generally, qualitative differences between SDMFT and MSMFT
may be found whenever degeneracies of the kind discussed
above appear in the ground states of the $\hat {\cal
K}^0_L(J=0)$ grand Hamiltonian. Such degeneracies appear in all
of the examples of loopholes considered
previously~\cite{Buonsante04a,Buonsante05a}. They also arise in
models dealing with two bosonic species~\cite{Chen10} and one
can expect differences between SDMFT and MSMFT to appear in
these cases as well. 

\section{Conclusions}
\label{sec:Conclusions}

In this paper we have developed, explained and utilized a mean-field 
theory of the Mott insulator-superfluid transition of bosons 
moving in optical lattices. In this approach, the lattice is 
partitioned into clusters which we refer to as
$L$-mers. In the Bose-Hubbard model the $L$-mers are coupled
by the hopping Hamiltonian; they can be isolated by invoking a
mean-field decoupling procedure whereby a superfluid order
parameter is introduced for all boundary sites of the $L$-mer.
This leads to a mean-field Hamiltonian taking the form shown in
Eq.~(\ref{KMF}). The resulting theory is thus a multi-site mean-field 
theory (MSMFT); it should be contrasted with site-decoupled 
mean-field theories which allow for different order 
parameters at different sites \cite{Chen10} but neglect
inter-site correlations.

The ground state of the mean-field Hamiltonian defines a grand
potential energy functional of the various order parameters.
The stationary points of this functional are shown to correspond
to self-consistent solutions in that the order parameters
evaluated at the stationary point coincide with the order
parameters defining the mean-field Hamiltonian itself. A
detailed analysis reveals that the stationary points are, in 
general, \textit{saddle points}. As a result, they cannot be
located by minimizing the energy functional. For 
weak hopping there is only one self-consistent solution, the Mott 
insulating phase, whereas at larger hopping a second stationary
point appears with lower energy, corresponding to the superfluid
phase. 

The identification of the phase boundaries separating the Mott 
insulating and superfluid phases can be analyzed using perturbation 
theory. Our work has clarified the relationship between two
different criteria for the determination of the phase
boundaries, one based on the energy functional itself and the
second based on the stability of an iterative map. We show that
the two criteria are equivalent and can be used interchangeably.

We have applied our theory to
the Bose-Hubbard model for $d$-dimensional hypercubic lattices, for 
$d=1$ (chains), $d=2$ (square lattice), and $d=3$ (simple cubic 
lattice). Our results demonstrate the improvements that MSMFT
affords relative to site-decoupled mean-field theories.
This, of course, is what one expects since the theory becomes
exact in the limit of an infinite $L$-mer. 
Specifically, our numerical results make clear 
that as the size of the $L$-mer \textit{and} the dimensionality of 
the system are increased, better agreement between MSMFT 
and numerical Monte Carlo data is obtained.
However, mean-field theories typically
underestimate the stability of the Mott insulating phase, and 
therefore (for a given chemical potential) underestimate the critical 
hopping at which this phase becomes unstable with respect to the 
superfluid phase.

In addition, we have applied our MSMFT to the analysis of
one-dimensional superlattices. For the dimer chain, the 
inequivalence of the two sites
within the dimer necessitates the introduction of two order
parameters. The stationary points of the resulting energy
functional are found to be saddle points in this case.
Apart from the underestimation of the critical hopping mentioned
above, the phase boundaries we obtain are in qualitative
agreement with those obtained on the basis of Monte Carlo
simulations~\cite{Rousseau06}. We also considered one example of
a more complex superlattice for which the predictions of SDMFT
and MSMFT differ qualitatively. Specifically, the phase diagram
as determined using MSMFT exhibits loopholes which are absent
in SDMFT. Such qualitative differences are expected to arise
when degeneracies exist in the $\hat {\cal K}^0_L(J=0)$ ground
state energy.

The MSMFT was used previously in a study of the disordered 
Bose-Hubbard model~\cite{Pisarski11}, and will be applied to 
other, more complicated situations in subsequent work.

\section*{APPENDIX}

A practical limitation of the MSMFT is the size of the Hilbert
space needed to diagonalize the mean-field Hamiltonians 
$\hat {\cal K}^0_L$ and $\hat {\cal K}^{MF}$ which grows
exponentially with the size of the $L$-mer. A natural basis is
provided by the occupation number states $|\{n_l\}\rangle$ where
$l=1,\dots,L$ and $n_l = 0,1,\dots,\infty$. For a given average
number of particles per site, it is sufficient to truncate the
range of $n_l$ at some maximum number $n_{max}$ which can be
varied to ensure convergence of the eigenstate calculations.
Within this truncated basis, the dimension of the basis is $S^L$
where $S=n_{max} +1$.

The perturbative calculations require the ground state of $\hat
{\cal K}^0_L$ for $N$ particles and excited states for
$N\pm1$ particles.
The truncated basis discussed above is excessive since it 
includes states with different total particle numbers. To obtain
the eigenstates within an $N$-particle subspace we therefore
retain only those states with the required number of
particles. We denote the states in this subset as $|k\rangle$,
$k = 1,\dots,N_{st}$, where $N_{st}$ is the number of states in
the set. These states are the occupation number states satisfying
$\sum_{l=1}^L n_l = N$ with $n_l \le n_{max}$. 

We now discuss how the size of the basis set can be reduced
further with the use of group theory~\cite{Fano92}. To be specific, we
consider the $3\times 3$ $L$-mer illustrated in
Fig.~\ref{fig:3x3}(b). The
symmetry operations which leave the $L$-mer invariant define a
group ${\cal G}$ of order $g$ ($g=8$ in this case) with group
elements $G_m$ consisting of the identity, three rotations and
four reflections. The savings provided by group theory stem
from the fact that the Hamiltonian $\hat {\cal K}^0_L$ commutes
with the group elements $G_m$ and that the ground state 
$|0\rangle$ of the $L$-mer
belongs to the identity irreducible representation with the
property $G_m|0\rangle = |0\rangle$. The reduced energy matrix
in (\ref{W_red_elem}) involves matrix elements of the form 
$\langle \nu |
\sum_{l \in A}(\hat a_l^\dagger + \hat a_l) |0\rangle$.
We observe that
\begin{eqnarray}
&&G_m \sum_{l\in A}\left(\hat{a}_l^{\dagger}+\hat{a}_l\right)
|0\rangle 
=  G_{m}\sum_{l\in A}\left(\hat{a}_l^{\dagger}+
\hat{a}_l\right)G_{m}^{-1}G_{m}|0\rangle \nonumber \\
&& \hskip 1.25truein = G_{m}\sum_{l\in A}\left(\hat{a}_l^{\dagger}+
\hat{a}_l\right)G_{m}^{-1}|0\rangle \nonumber \\
&& \hskip 1.25truein = \sum_{l\in A}\left(\hat{a}_l^{\dagger}+
\hat{a}_l\right)|0\rangle.
\end{eqnarray}
The last step is a consequence of the invariance of the sum of field
operators under a symmetry operation. The state $\sum_{l\in
A}\left(\hat{a}_l^{\dagger}+\hat{a}_l\right) |0\rangle$ thus
belongs to the identity representation and as a result, the state
$|\nu \rangle$ must also belong to this representation for the
matrix element to be finite. It is clear from this discussion
that the calculation of the matrix in (\ref{C}) cannot be
simplified using group theory since the action of $\hat c_\alpha$ on
$|0\rangle$ creates a state that belongs to other irreducible
representations.

Since $|\nu\rangle$ must belong to the identity representation,
it can be expanded in terms of states belonging to this
representation. These states are constructed as follows,
\begin{equation}\label{symmstate}
|\kappa\rangle=\frac{1}{\sqrt{N_{\kappa}}}\sum_{m=1}^{g}G_{m}|k\rangle,
\end{equation}
where $1/\sqrt{N_\kappa}$ is an appropriate normalization
constant. The number of distinct states formed in this way is
$N_{sym} < N_{st}$. This reduction in the dimension of the basis
set is the advantage afforded by the use of group theory.

Within this symmetrized basis, an eigenstate of $\hat{\mathcal
K}^{0}$ is expanded as
\begin{equation}\label{symeigenvec}
|\nu\rangle=\sum_{\kappa=1}^{N_{sym}}c^{(\nu)}_{\kappa}|\kappa\rangle.
\end{equation}
The expansion coefficients and corresponding eigenvalues
$\Omega_\nu(\{{\bf 0}\})$ are then obtained by the
diagonalization of the Hamiltonian matrix $\langle \kappa
|\hat{\mathcal K}^{0}_L |\kappa'\rangle$. The reduction of the 
dimension of the eigenvalue problem from $N_{st}$ to $N_{sym}$
is roughly a factor $g$, the order of the group, and represents
a significant computational saving. However, even with the use
of group theory, the $4\times 4$ $L$-mer for a 2D square lattice
remains a formidable calculation.

\acknowledgments
This work was supported by grants from the Natural Sciences and
Engineering Research Council of Canada. We would also like to
thank Vittoria Penna for useful discussions.


\begin{thebibliography}{99}
\bibitem{footnote} SDMFT for a homogenous lattice is recovered
by noting that the index $\alpha$ takes only one value and that 
$g= z$, the number of nearest neighbours in the lattice.
\bibitem{footnote1}
The parameters are obtained by a
nonlinear fit of $\ln \bar C_l(L)$ to the logarithm of
(\ref{eqn:fits}).
\end{thebibliography}
\end{document}